\documentclass[aps,prd,amsmath,amssymb, preprintnumbers, reprint,longbibliography,superscriptaddress,nofootinbib]{revtex4-1}
\pdfoutput=1
\usepackage{lmodern, graphicx, multirow, xcolor, adjustbox}
\usepackage[colorlinks=true, citecolor=blue, urlcolor=blue, linkcolor=blue, breaklinks=true, pdfpagelabels=false]{hyperref}

\makeatletter\g@addto@macro\bfseries{\boldmath}\makeatother
\DeclareMathSymbol{\shortminus}{\mathbin}{AMSa}{"39}

\catcode`\$=\active
\gdef$#1${\texorpdfstring{\(#1\)}{\detokenize{#1}}}

\usepackage{wasysym}
\usepackage{lipsum}
\usepackage{multirow}
\usepackage{verbatim}
\usepackage{amsmath}

\usepackage{cancel}
\usepackage{hyperref}
\usepackage{makecell}
\usepackage{float}

\usepackage{pifont}

\newcommand{\mm}[1]{\textcolor{red!90!black}{MM[#1]}}

\begin{document}

\newcommand{\cpthree}{Center for Cosmology, Particle Physics and Phenomenology (CP3), Universit\'{e} Catholique de Louvain, 1348 Louvain-la-Neuve, Belgium}

\author{C\'eline Degrande}
\email{celine.degrande@uclouvain.be}
\affiliation{\cpthree}

\author{Matteo Maltoni}
\email{matteo.maltoni@uclouvain.be}
\affiliation{\cpthree}

\preprint{IRMP-CP3-25-36}

\title{Constraining the four-light quark operators in the SMEFT with multijet and VBF processes at linear level}

\begin{abstract}
We investigate how the interference of the SM with ten four-light quark operators in the SMEFT can be constrained thanks to multijet and $Z$, $W$, $\gamma$ VBF production in association with jets. The differential distributions for each process are generated at LO for different jet multiplicities, that are then merged and showered. We check which observables provide better bounds on the Wilson coefficients, and what directions in the ten-dimensional coefficient space they are able to probe. We discuss the relevance of the quadratic contributions with respect to the linear terms and use them to assess the validity of the EFT approach.
\end{abstract}

\maketitle

\subparagraph{Introduction}
With the current measurements agreeing with the Standard Model (SM), one possibility is that new resonances, if any, might exist at high energies that cannot be reached by current and future accelerators. The Standard Model Effective Field Theory (SMEFT) is a formalism to parametrise eventual deviations from the SM predictions, induced by the interactions among known states and the new heavy ones, at the energies of current experiments. Complete sets of higher-dimensional effective operators $O_i$, multiplied by their Wilson coefficients $C_i$, are added to the SM Lagrangian,
\begin{equation}
   \mathcal{L}_{\small{SMEFT}} = \mathcal{L}_{\small{SM}} + \sum_{i} \frac{C_i}{\Lambda^2}\hspace{1mm}O_i + \mathcal{O}(1/\Lambda^4),
\end{equation}
with $\Lambda$ a cutoff scale. This Lagrangian can describe any new physics (NP) which is heavy enough compared to the considered experiment. This criterion assesses the validity of the effective approach, since the truncation at the order $\mathcal{O}(1/\Lambda^2)$ provides a good approximation only if the higher-order terms can be neglected. It is however hard to check this in practice, as there is no direct access to the value of $\Lambda$ independently of $C_i$. The same expansion is observed at differential level for a generic measurable variable $X$,
\begin{equation}
   \frac{d\sigma}{dX} = \frac{d\sigma^{SM}}{dX} + \sum_i  \frac{C_i}{\Lambda^2}\frac{d\sigma^{1/\Lambda^2}}{dX}+ \mathcal{O}(1/\Lambda^4), \label{smeft_xs}
\end{equation}
where the second term, of order $\mathcal{O}(1/\Lambda^2)$ in the expansion, is an interference between the SM and the dimension-6 operators. The last term contains the cross section from the diagrams with the insertion of a dimension-6 operator squared, plus the interference of the SM with dimension-8 operators. Due to the large number of the latter \cite{dim8_basis,dim8_basis_1}, the dimension-6 squared is usually considered as an approximation for the missing terms in the truncated series. It can be used as a practical error estimate for the Effective-Field-Theory (EFT) expansion and, with a bit of care\footnote{When the interference is suppressed, this term can be larger than the interference even if the EFT expansion is still valid}, as a test of the EFT validity since it should be smaller than the interference, due to the higher suppression by the cutoff scale. The validity of the EFT is an important subject that triggered a lot of discussion \cite{Heinrich_2022,brivio2022,Allwicher_2025}.

Among the operators that have been included in the Warsaw basis for dimension 6 \cite{Grzadkowski:2010es,Buchmuller:1985jz}, the four-light quark ones (4LQ) introduce four-fermion contact interactions that are not present in the SM, featuring the $u,d,c,s,b$ quarks and their antiquarks. At Leading Order (LO), these operators do not contribute to the main top and Higgs processes and thus they have been investigated less and almost never included in global fits with top, electroweak (EW) and flavour data, as NP is often expected to couple preferentially to the heaviest SM states \cite{globalfits_1, globalfits_2, globalfits_3, globalfits_4, globalfits_5, globalfits_6,isidori,Wang:2024zns}. While this assumption is a model-builder bias, the top and Higgs interactions have been among the least constrained since they have been discovered more recently, which has further contributed to the exploration of their properties.

However, the 4LQ operators can induce corrections to virtually any process at Next-to-Leading Order (NLO), if two of the fermionic lines in the diagrams they produce are closed into a loop. In particular, they have been probed through the running amongst the operators in top, EW and flavour data, that offer complementary constraints to those targeted in this paper. 
Like any four-fermion operators, the 4LQ ones can be generated easily at tree-level through the exchange of a heavy resonance (see {\it e.g.} \cite{deBlas:2017xtg}). Therefore, they may have a similar effect, even at one-loop, to operators that contribute at tree-level but are generated by the UV model at the loop level ({\it e.g.} the operators with three strength tensors).

We consider ten 4LQ operators, defined in Table \ref{tab:ops_def}; they conserve CP and the lepton and baryon numbers. The aim of this letter is to test how currently-measured observables and processes can be used to set bounds on their Wilson coefficients. It is known that dijets can probe at most two directions in this coefficient space \cite{2j_SMEFT}, so we also check processes where jets are produced together with an EW boson: since the latter can be sensitive to the quantum numbers of the various quark fields, this can help probe additional directions.
In particular, we focus on multijet production and $\gamma +$jets, $Z+$jets and $W+$jets Vector-Boson Fusion (VBF) processes, as they receive contributions from the considered operators at LO. Additionally, we exploit the most recent developments in flavour-tagging of $b$- and $c$-jets to be more sensitive to some subprocesses.

A notebook with all the distributions and $\chi^2$ functions employed to obtain the limits presented in this study can be found linked to this letter.

\begin{table} 
\caption{\small{List of 4LQ operators considered in this study. $q$ denotes the left-handed quark doublets of all three generations, while $u,d$ are the right-handed up-type and down-type quark fields, respectively. $\sigma^I$, with $I=\{1,2,3\}$, are the Pauli matrices, while $T^A$, $A=\{1,\ldots,8\}$ are the $SU(3)_C$ generators. $p,r,s,t$ are flavour indices (see Sec. \ref{sec:framework} for details); the spin and colour ones are omitted. For each process, $\checkmark$ and $\times$ specify if the interference of an operator with the SM QCD contributes to it or not}} \label{tab:ops_def}
\begin{tabular}{c|cc|cccc}
   \hline
   Operator & Coeff. & Definition & $jj$ & $Zjj$ & $Wjj$ & $\gamma jj$ \\
   \hline
   $O_{qq}^{(1)}$ & $C_{qq}^{(1)}$ & $(\bar{q}_p \gamma^\mu q_r)(\bar{q}_s \gamma_{\mu} q_t)$ & $\checkmark$ & $\checkmark$ & $\checkmark$ & $\checkmark$ \\
   $O_{qq}^{(3)}$ & $C_{qq}^{(3)}$ & $(\bar{q}_p \gamma^\mu \sigma^I q_r)(\bar{q}_s \gamma_{\mu} \sigma^I q_t)$ & $\checkmark$ & $\checkmark$ & $\checkmark$ & $\checkmark$ \\
   $O_{uu}$ & $C_{uu}$ & $(\bar{u}_p \gamma^\mu u_r)(\bar{u}_s \gamma_{\mu} u_t)$ & $\checkmark$ & $\checkmark$ & $\times$ & $\checkmark$ \\
   $O_{dd}$ & $C_{dd}$ & $(\bar{d}_p \gamma^\mu d_r)(\bar{d}_s \gamma_{\mu} d_t)$ & $\checkmark$ & $\checkmark$ & $\times$ & $\checkmark$ \\
   $O_{ud}^{(1)}$ & $C_{ud}^{(1)}$ & $(\bar{u}_p \gamma^\mu u_p)(\bar{d}_s \gamma_{\mu} d_s)$ & $\times$ & $\checkmark$ & $\times$ & $\checkmark$ \\
   $O_{ud}^{(8)}$ & $C_{ud}^{(8)}$ & $(\bar{u}_p \gamma^\mu T^A u_p)(\bar{d}_s \gamma_{\mu} T^A d_s)$ & $\checkmark$ & $\checkmark$ & $\times$ & $\checkmark$ \\
   $O_{qu}^{(1)}$ & $C_{qu}^{(1)}$ & $(\bar{q}_p \gamma^\mu q_p)(\bar{u}_s \gamma_{\mu} u_s)$ & $\times$ & $\checkmark$ & $\checkmark$ & $\checkmark$ \\
   $O_{qu}^{(8)}$ & $C_{qu}^{(8)}$ & $(\bar{q}_p \gamma^\mu T^A q_p)(\bar{u}_s \gamma_{\mu} T^A u_s)$ & $\checkmark$ & $\checkmark$ & $\checkmark$ & $\checkmark$ \\
   $O_{qd}^{(1)}$ & $C_{qd}^{(1)}$ & $(\bar{q}_p \gamma^\mu q_p)(\bar{d}_s \gamma_{\mu} d_s)$ & $\times$ & $\checkmark$ & $\checkmark$ & $\checkmark$ \\
   $O_{qd}^{(8)}$ & $C_{qd}^{(8)}$ & $(\bar{q}_p \gamma^\mu T^A q_p)(\bar{d}_s \gamma_{\mu} T^A d_s)$ & $\checkmark$ & $\checkmark$ & $\checkmark$ & $\checkmark$ \\
   \hline
\end{tabular}
\end{table}

\section{\label{sec:framework}Framework}
Our analysis is performed via \textsc{MadGraph5}\_a{\sc MC@NLO} v3.5.4 \cite{Alwall:2014}, but we advise to always use the latest version, as issues were fixed during this work. We feed it a Universal FeynRules Output ({\sc UFO}) \cite{Degrande:2011ua}, written from a FeynRules model \cite{Alloul:2013bka} that contains the SM and the ten 4LQ operators in Table \ref{tab:ops_def}\footnote{See https://feynrules.irmp.ucl.ac.be/wiki/4LQ}. The CKM matrix is assumed to be diagonal. The leptons and quarks, except the top, are considered as massless. 

The top quark is included in the model, but not in the process definitions. In principle, top and light-quark operators should be separated, but since the top processes are not taken into account, the top quark does not have an effect in this work. 
The effect of the bottom quark is also limited, as it is quite suppressed by the Parton Density Function (PDF) in the initial state. It can increase the SMEFT contribution when it appears in the final state, but this is not expected to have a large impact as the contributions from the bottom either are small or have the same shapes as the others. The only exception comes from processes where $b$-tagging is employed, but their constraining power seems to be, as it will be shown, not competitive with the others. Therefore, our scenario is relatively optimistic.
In a minimal-flavour-violation scenario where only the top is massive, the operators featuring the top should be separated from the ones with all the other flavours and this would further increase the parameter space, but they would not contribute to the processes studied here.

In all the non-SM distributions and cross sections presented in this paper, $C_i/\Lambda^2$ is set to 1 TeV${}^{-2}$ for each operator.

For all processes, we include diagrams with both QED and QCD vertices and generate up to three jets at LO parton level, using {\sc MLM} for matching and merging \cite{mlm} and then showering the events with {\sc Pythia8} \cite{Pythia:2015}. We expect the largest corrections to the partonic LO results to be due to the matching and merging and the parton shower (PS), as we focus on differential measurements of processes with jets.
The sum of transverse energies divided by two, $H_T/2$, is chosen as the dynamical scale for the events. First, we reconstruct eventual dressed leptons with the $k_t$ algorithm and a radius parameter $R=0.1$ \cite{kt_1,kt_2,Fastjet:2012}; then, jets are obtained from the remaining final states using the anti-$k_t$ algorithm \cite{AntiKt:2008} with $R=0.4$. Predictions are presented for the Large Hadron Collider (LHC) at 13 TeV.

Numerical and scale uncertainties are reported for each result \cite{Frederix:2012}. Numerical errors are due to the limited number of events generated, while scale variations are computed by taking the envelope of nine scale combinations, in which the renormalisation and factorisation scales $\mu_{R,F}$ are varied by factors 0.5 and 2.
Even if large samples are generated at the partonic level, only a faction of events survive after the shower and the cuts, especially in the tails of the distributions where uncertainties can be significant. While the numerical errors are in general small (or at least smaller than the other ones), they should be kept in mind when constraining the subleading directions in the coefficient space, which rely on shape variations between distributions from different operators: these are slight and can be dominated by the numerical uncertainties. Indeed, even if they are almost uncorrelated between bins, they can induce small shape differences of pure numerical origin among the operators. The PDF errors are not reported, but were found to be smaller than the scale ones by at least a factor $\sim 3$ in every distribution bin for multijet production.

We leave the study of renormalisation-group evolution (RGE) for future work, as our main focus is to look for the operator effects on distribution shapes, counting the number of constraints and quantifying their strength. Once included, RGEs would allow to obtain bounds from flavour observables as, even if the 4LQ operators are flavour-conserving, some violation still occurs in the SM \cite{matching_flav_symm}. Moreover, when the RGE mixing effects are considered, Drell-Yan processes can be used to constrain these 4LQ operators as well \cite{miralles}.

\subparagraph{Flavour-indices contractions}
For operators featuring two fermionic currents with same chirality, like $O_{qq}^{(1)} = (\bar q_p \gamma^\mu q_r)(\bar q_s \gamma_\mu q_t)$, two contractions of the generation indices $p,r,s,t$ are possible: inside the fermion bilinears ($\delta_{pr} \delta_{st}$) or between them ($\delta_{pt} \delta_{rs}$). The same happens for $O_{qq}^{(3)}$, $O_{uu}$ and $O_{dd}$. In our model, these two cases are summed together inside the same operator: therefore they contain both the colour-singlet and octet terms and interfere with both gluon and EW-boson diagrams. The contribution with four equal flavour indices, being identical in the two contractions, is however added only once. Since we want to keep the analysis simple and estimate the best bounds that can be placed over these objects, we leave the separation of these components for future studies\footnote{This would be needed also for the RGE, as the octet and singlet components run separately}; this would increase the parameter space and probably yield looser limits than the ones presented in this letter. 
For all the others operators, only the $\delta_{pr} \delta_{st}$ contraction is considered, as shown in Table \ref{tab:ops_def}.

\subparagraph{Constrained directions}
For a generic observable $X$, a $\chi^2$ function is built from the experimental, SM and interference differential distributions, as
\begin{equation}
   \chi^2 = \sum\limits_{i=1}^{N_\text{bins}} \frac{1}{\Delta_i^2} \Bigg( \frac{d\sigma^\text{exp}_i}{dX}-\frac{d\sigma^\text{SM}_i}{dX} -\sum\limits_j^{\scriptscriptstyle \text{operators}} \frac{c_j}{\Lambda^2} \frac{d\sigma^{1/\Lambda^2}_{j,i}}{dX} \Bigg)^2, \label{chisq}   
\end{equation}
where $\Delta_i$ is the sum in quadrature of the numerical and scale uncertainties from the experimental and SM results in the $i^{th}$ bin and $\mathbf{c}^T = \left( C_{qq}^{(1)}, C_{qq}^{(3)}, \ldots, C_{qd}^{(8)} \right)$ is the vector of 4LQ coefficients, so that the second sum runs over all the operators.

In order to get an estimate of how the limits would change when uncertainties on the interference are considered, we compare the results from the previous formula against the ones from
\begin{multline}
   \chi^2 = \sum\limits_{i=1}^{N_\text{bins}} \frac{1}{\Delta_i^2} \Bigg[ \frac{d\sigma^\text{exp}_i}{dX}-\frac{d\sigma^\text{SM}_i}{dX}+ \\
   -\sum\limits_j^{\scriptstyle \text{operators}} \frac{c_j}{\Lambda^2} \Bigg( \frac{d\sigma^{1/\Lambda^2}_{j,i}}{dX} \pm \Delta^{1/\Lambda^2}_{j,i} \Bigg) \Bigg]^2, \label{chisq_2}   
\end{multline}
where $\Delta^{1/\Lambda^2}_i$ contains the theoretical errors on each interference differential distribution, {\it i.e.} the numerical and scale-variation ones. This includes the correlations between bins, as the same scale is chosen for all of them. Among the possibilities that are obtained when adding or subtracting these terms, the one that gives the largest bounds is considered. For the distributions generated by us, numerical uncertainties are summed in quadrature, while scale variations are combined linearly to partially account for their correlations between all the operators \cite{2j_tevatron}. Since the full correlation matrices are not available for all the processes, only the limits from Eq. \eqref{chisq} are reported, while the ones from Eq. \eqref{chisq_2} are used for comparison.
For each operator, the individual bounds are obtained by setting all the other coefficients to 0 in the $\chi^2$ formulas above, while for the marginalised limits they are set to the values that minimise the $\chi^2$ function along their directions.

The vector of uncorrelated directions that can be constrained by the measurements is named $\mathbf{c^\prime}^T = \big( C_1,C_2,\ldots,C_{10} \big)$ and is related to $\mathbf{c}$ through a change of basis
\begin{equation}
   \mathbf{c^\prime} = \mathcal{R}^T \cdot \mathbf{c}.
\end{equation}
Since Eqs. \eqref{chisq} and \eqref{chisq_2} are quadratic polynomials in the Wilson coefficients, the columns of $\mathcal{R}$ are the eigenvectors of the matrix of coefficients of the quadratic terms in the $\chi^2$ expression \cite{andres_heavyquarks}. These eigenvectors represent the directions, in the coefficients space, that can be constrained by the data: the larger the eigenvalues associated with them, the better they can be probed. Some among them are flat directions and, in principle, they should be related to null eigenvalues; because of the numerical nature of the computation, though, it might be tricky to identify them. To do so, we estimate the uncertainty of each eigenvalue as in \cite{ew_4f_smeft}: from each distribution, we generate numerous toy ones by replacing each bin content with a random number, extracted from a Gaussian centred on that bin value and with standard deviation equal to the Monte Carlo (MC) uncertainty. The eigenvalues are then computed from each set of toy plots and their standard deviation $\sigma$ is taken as the error. Each eigenvalue that is compatible with 0 within $2\sigma$ is considered to be related to a flat direction. We define the eigenvectors so that their norm is unitary and they show a positive scalar product with the unit vector $(1,1,\ldots,1)$.

\subparagraph{Squared contributions}
Even if the $\mathcal{O}(1/\Lambda^4)$ term in Eq. \eqref{smeft_xs} is usually subleading to the interference one at the experimental energies, it might dominate the cross-section deviation from the SM at the high energies that are considered in this analysis, depending on the constraints and if the interference between the SM and the dimension-6 operators is suppressed. 
Its inclusion would require the computation of the diagrams that feature the insertion of a dimension-6 operator squared, with the incorporation of the correlations among them, and of the interference between the SM and the dimension-8 operators, whose number is larger than the dimension-6 one. 
In order to get an estimate of this effect without complicating the study, we compute the cross section of the $O_{qq}^{(3)}$ quadratic contribution to each process, and compare it with the linear one for coefficient values that are close to the limits we obtain.

The interference suppression is not present for multijet production, as it can be estimated from Table~\ref{tab:all_xsects} using the Schwartz bound and expected from helicities selection rules~\cite{Azatov:2016sqh}. Because of this strong theoretical argument, the same is true for the other operators as well. We also check at the partonic level that the ratio of the interference cross section and Schwartz bound is of order $\sim 1$, except if there is a suppression due to the colour such as for $O_{ud}^{(1)}$ in dijet production. As a result, for coefficient values close to the limits we obtain, the dimension-6 square terms provide good estimates of the $\mathcal{O}(1/\Lambda^4)$ contributions. 

\subparagraph{Jet tagging}
Flavour-tagging can increase the sensitivity to operators that feature only up- or down-like quarks. An example of the application of such algorithms to $Z+$jets production in the SM can be found in \cite{zjj_flavour_tag}.
For multijet production, we emulate the DL1r algorithm \cite{bc_tagging} for $b$- and $c$-tagging. Reconstructed jets from \textsc{FastJet}, if they are within a $\Delta R= 0.4$ radius from a MC-true $b$-jet, are $b$-tagged with a $p_T$-dependent efficiency, whose values are digitised from Figures 11 and 12 in the reference paper, respectively for the 20-250 and 250-3000 GeV transverse momentum ranges in the 77\% working point. Light-flavour and $c$-jets are mistagged as $b$-ones according to the rejection rates shown in the same figures. The dependencies on pseudorapidity and luminosity are not considered. For the $c$-tagging, the same procedure is followed, but the efficiency is fixed at 30\% independently of the $p_T$, and the $b$- and light-flavour jets mistagging rates are also constant and taken from Fig. 16 of the reference for the 30\% working point. The $c$-tagging procedure is applied only to jets with a $p_T$ between 20 and 250 GeV. Since we do not have the full output of the DL1r network, if a jet is $b$- and $c$-tagged at the same time, the MC truth is consulted, or a random extraction is applied in case of doubly-mistagged light jets.

\begin{table*} 
\caption{\small{Cross sections for the contributions of the SM and its interference with the 4LQ operators, whose coefficients are set to 1 TeV${}^{-2}$, to the processes investigated in this analysis. The dim.-6 squared contributions to the $\mathcal{O}(1/\Lambda^4)$ term for $O_{qq}^{(3)}$ are also shown. The units are not the same in each column and can be found at the top. The 4LQ cross sections are computed at LO with MLM and PS, and the first relative uncertainty is numerical, while the following numbers are scale variations. For the SM, we report the values from the respective experimental analyses for $Z+$jets and $\gamma +$jets, and the ones we obtain in our LO generations for $W+$jets and multijets. The experimentally-measured cross sections are also shown, when available, with their cumulative uncertainties}} \label{tab:all_xsects}
\resizebox{\textwidth}{!}{
\begin{tabular}{c|ccccc}
   \hline
    & \multicolumn{2}{c}{Multijets} & $\ell^+ \ell^- +$jets & $\ell^\pm \nu +$jets & $\gamma +$jets \\
    & $M_{jj}>2.4$ TeV (pb) & $M_{jj}>6$ TeV {\bf (fb)} & {\bf (fb)} & (pb) & {\bf (fb)} \\ \hline \hline
   Exp. & Not avail. & Not avail. & 3.7$\cdot 10^1 {}^{+11\%}_{-11\%}$ & Not avail. & 2.3$\cdot 10^4 {}^{+5\%}_{-5\%}$ \\
   SM & 6.6$\pm 0.6\%^{+15\%}_{-29\%}$ & 2.7$\pm 0.5\%^{+26\%}_{-30\%}$ & 3.95$\cdot 10^1 {}^{+9\%}_{-9\%}$ & 4.2$\cdot 10^1\pm 0.5\%^{+10\%}_{-17\%}$ & 2.6$\cdot 10^4 {}^{+31\%}_{-31\%}$ \\ \hline
   & \multicolumn{5}{c}{$\mathcal{O}(1/\Lambda^2)$} \\ 
   $O_{qq}^{(1)}$ & -5.9$\pm 0.7\%^{+27\%}_{-17\%}$ & -3.7$\cdot 10^1 \pm 0.5\%^{+35\%}_{-27\%}$ & -5.12$\pm 0.8\%^{+21\%}_{-14\%}$ & -1.3$\pm 0.8\%^{+31\%}_{-23\%}$ & -4.0$\cdot 10^2 \pm 0.3\%^{+13\%}_{-18\%}$ \\
   $O_{qq}^{(3)}$ & -1.8$\cdot 10^1 \pm 0.6\%^{+22\%}_{-11\%}$ & -7$\cdot 10^1 \pm 0.4\%^{+43\%}_{-23\%}$ & -3.7$\cdot 10^1\pm 0.5\%^{+16\%}_{-11\%}$ & -6.1$\pm 0.5\%^{+18\%}_{-11\%}$ & -1.9$\cdot 10^3\pm 0.15\%^{+9\%}_{-11\%}$ \\
   $O_{uu}$ & -4.4$\pm 0.5\%^{+16\%}_{-14\%}$ & -2.7$\cdot 10^1 \pm0.4\%^{+37\%}_{-26\%}$ & -3.8$\cdot 10^{-1}\pm 0.5\%^{+18\%}_{-8\%}$ & $\times$ & -3.7$\cdot 10^2 \pm 0.16\%^{+15\%}_{-14\%}$ \\
   $O_{dd}$ & -9.4$\cdot 10^{-1} \pm 0.5\%^{+16\%}_{-14\%}$ & -1.7$\pm 0.5\%^{+41\%}_{-24\%}$ & -3.3$\cdot 10^{-2}\pm 0.5\%^{+18\%}_{-9\%}$ & $\times$ & -3.9$\cdot 10^1 \pm 0.18\%^{+10\%}_{-13\%}$ \\
   $O_{ud}^{(1)}$ & $\times$ & $\times$ & 1.3$\cdot 10^{-2}\pm 0.6\%^{+15\%}_{-31\%}$ & $\times$ & 1.5$\cdot 10^1 \pm 0.2\%^{+27\%}_{-13\%}$ \\
   $O_{ud}^{(8)}$ & -1.03$\pm 0.6\%^{+16\%}_{-14\%}$ & -3.2$\pm 0.4\%^{+37\%}_{-25\%}$ & -1.02$\cdot 10^{-1}\pm 0.6\%^{+22\%}_{-14\%}$ & $\times$ & -1.13$\cdot 10^2 \pm0.18\%^{+8\%}_{-12\%}$ \\
   $O_{qu}^{(1)}$ & $\times$ & $\times$ & -5.6$\cdot 10^{-2}\pm 0.7\%^{+18\%}_{-13\%}$ & -1.1$\cdot 10^{-2}\pm 1.2\%^{+63\%}_{-46\%}$ & -1.3$\cdot 10^1 \pm 0.3\%^{+23\%}_{-31\%}$ \\
   $O_{qu}^{(8)}$ & -2.0$\pm 0.5\%^{+15\%}_{-15\%}$ & -6.9$\pm 0.4\%^{+12\%}_{-26\%}$ & -9.3$\cdot 10^{-1}\pm 0.5\%^{+20\%}_{-12\%}$ & -1.9$\cdot 10^{-1}\pm 0.5\%^{+21\%}_{-16\%}$ & -2.1$\cdot 10^2\pm 0.2\%^{+14\%}_{-24\%}$ \\
   $O_{qd}^{(1)}$ & $\times$ & $\times$ & 1.5$\cdot 10^{-2}\pm 1.3\%^{+20\%}_{-27\%}$ & 1.1$\cdot 10^{-3}\pm 9\%^{+270\%}_{-360\%}$ & 2.0$\pm 0.5\%^{+60\%}_{-60\%}$ \\
   $O_{qd}^{(8)}$ & -1.1$\pm 0.5\%^{+27\%}_{-15\%}$ & -2.4$\pm 0.5\%^{+42\%}_{-25\%}$ & -4.8$\cdot 10^{-1}\pm 0.6\%^{+21\%}_{-13\%}$ & -1.5$\cdot 10^{-1}\pm 0.5\%^{+27\%}_{-13\%}$ & -8.0$\cdot 10^1\pm 0.2\%^{+15\%}_{-24\%}$ \\
   \hline
   & \multicolumn{5}{c}{Partial $\mathcal{O}(1/\Lambda^4)$} \\ 
   $O_{qq}^{(3)}$ & 1.0$\cdot 10^2 \pm 0.9\%^{+20\%}_{-10\%}$ & 1.9$\cdot 10^3 \pm 0.6\%^{+48\%}_{-20\%}$ & 9.8$\cdot 10^1 \pm 1.2\%^{+10\%}_{-16\%}$ & 1.8$\cdot 10^1 \pm 0.6\%^{+7\%}_{-11\%}$ & 6.2$\cdot 10^3 \pm 0.4\%^{+8\%}_{-8\%}$ \\
   \hline
\end{tabular}
}
\end{table*}

\begin{figure}
   \caption{\small{Shapes of the differential cross section with respect to $\chi_{jj}$ for the SM and its interference with the 4LQ operators, for the dominant subprocess $uu \rightarrow uu$ to dijet production at LO. PDF and PS effects are not included}} \label{fig:dxsdChi}
   \includegraphics[width=0.49\textwidth]{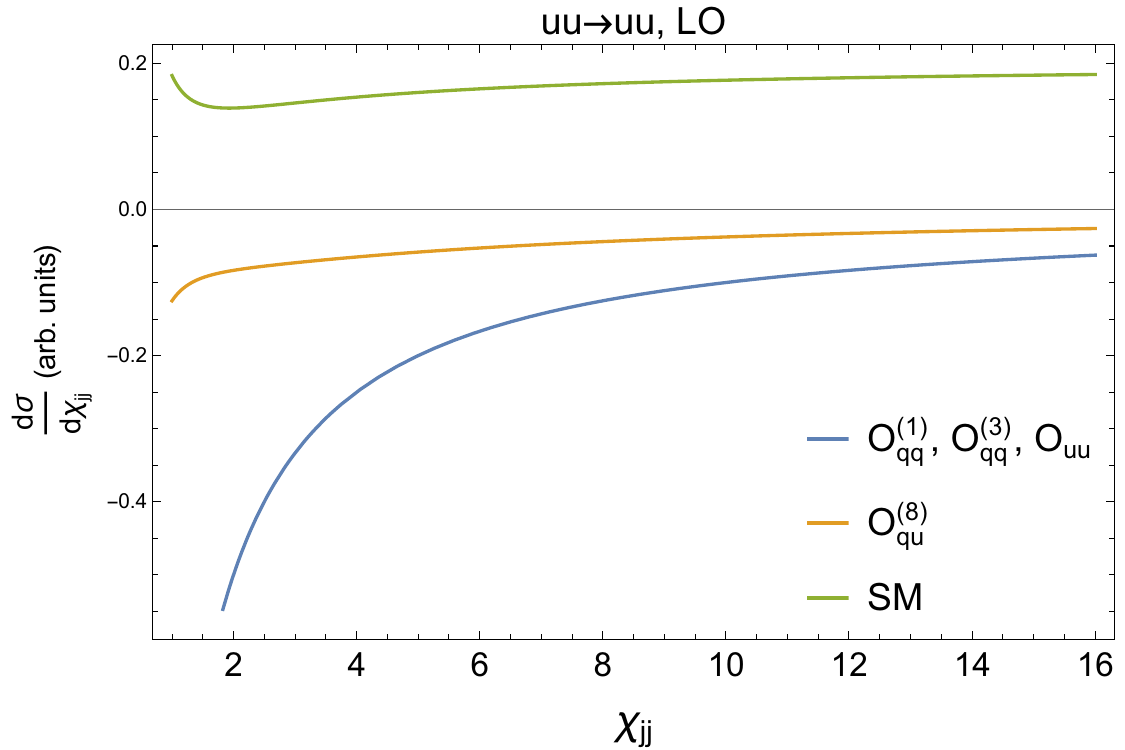}
\end{figure}

\begin{figure*}
   \caption{\small{Differential ({\it top}) and normalised ({\it bottom}) distributions of the exponential of the azimuthal distance among the two leading jets in multijet production, for the SM ({\it black}) and the contributing 4LQ operators, with all the coefficients set to $C_i/\Lambda^2 = 1$ TeV${}^{-2}$. Two dijet invariant-mass regions are shown: [2.4, 3] TeV ({\it left}) and [6, 13] TeV ({\it right}). The numerical uncertainties are represented with shaded bands in both plots, while the scale variations are shown in hatched bands in both plots for the SM, and only on top for the 4LQ operators. Note that the cross-section unit is not the same in the two top plots. The experimental measurements are also included in the bottom plots}} \label{fig:jj_chijj}
   \includegraphics[width=.49\textwidth]{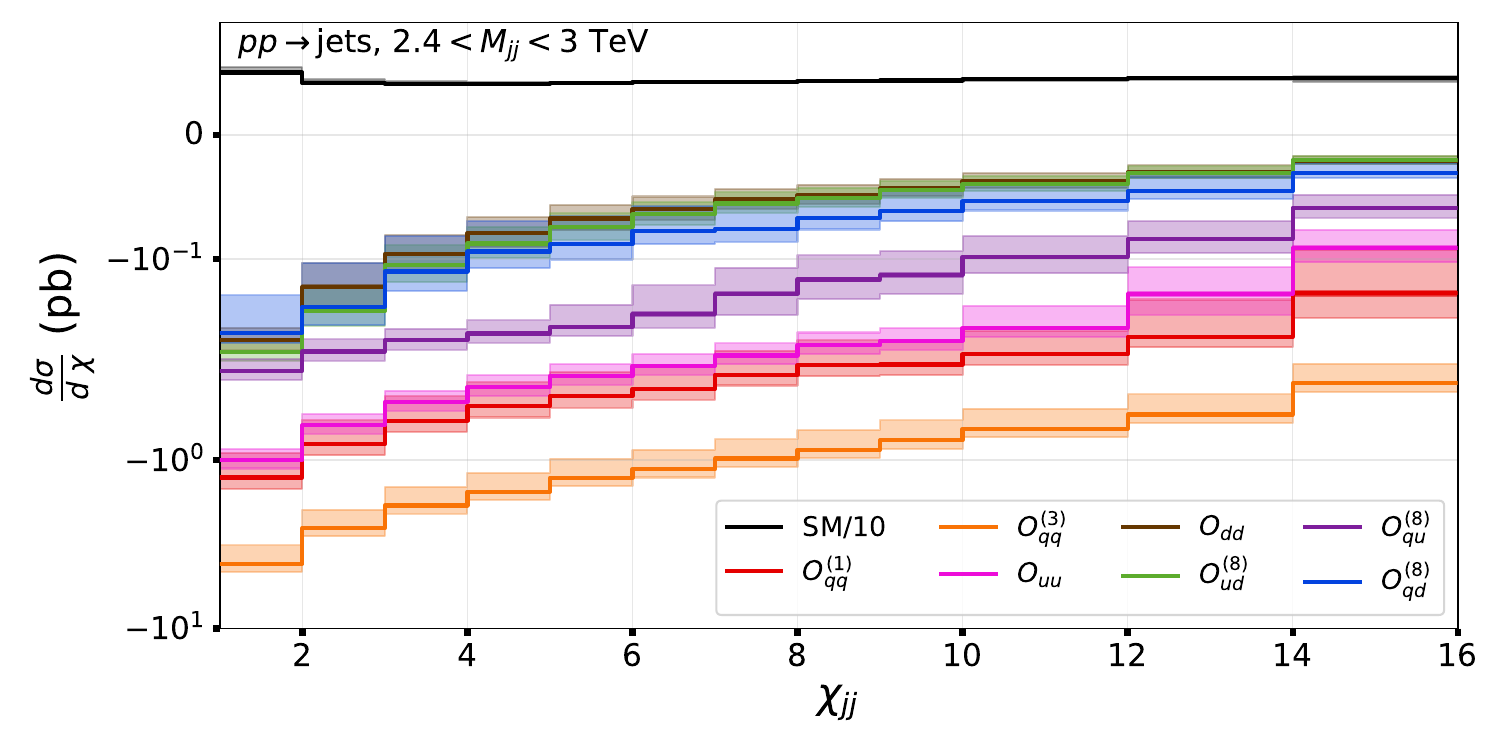}
   \includegraphics[width=.49\textwidth]{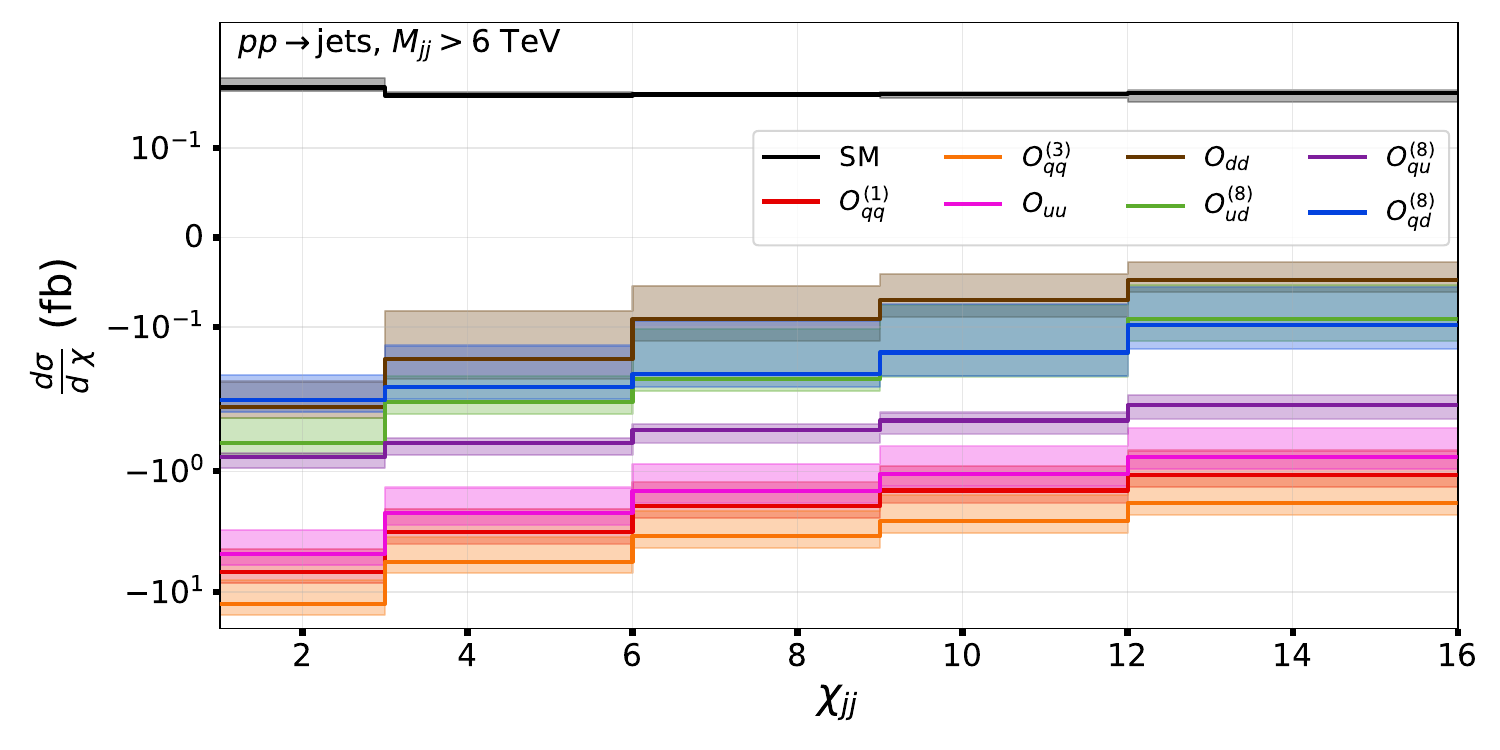}
   \includegraphics[width=.49\textwidth]{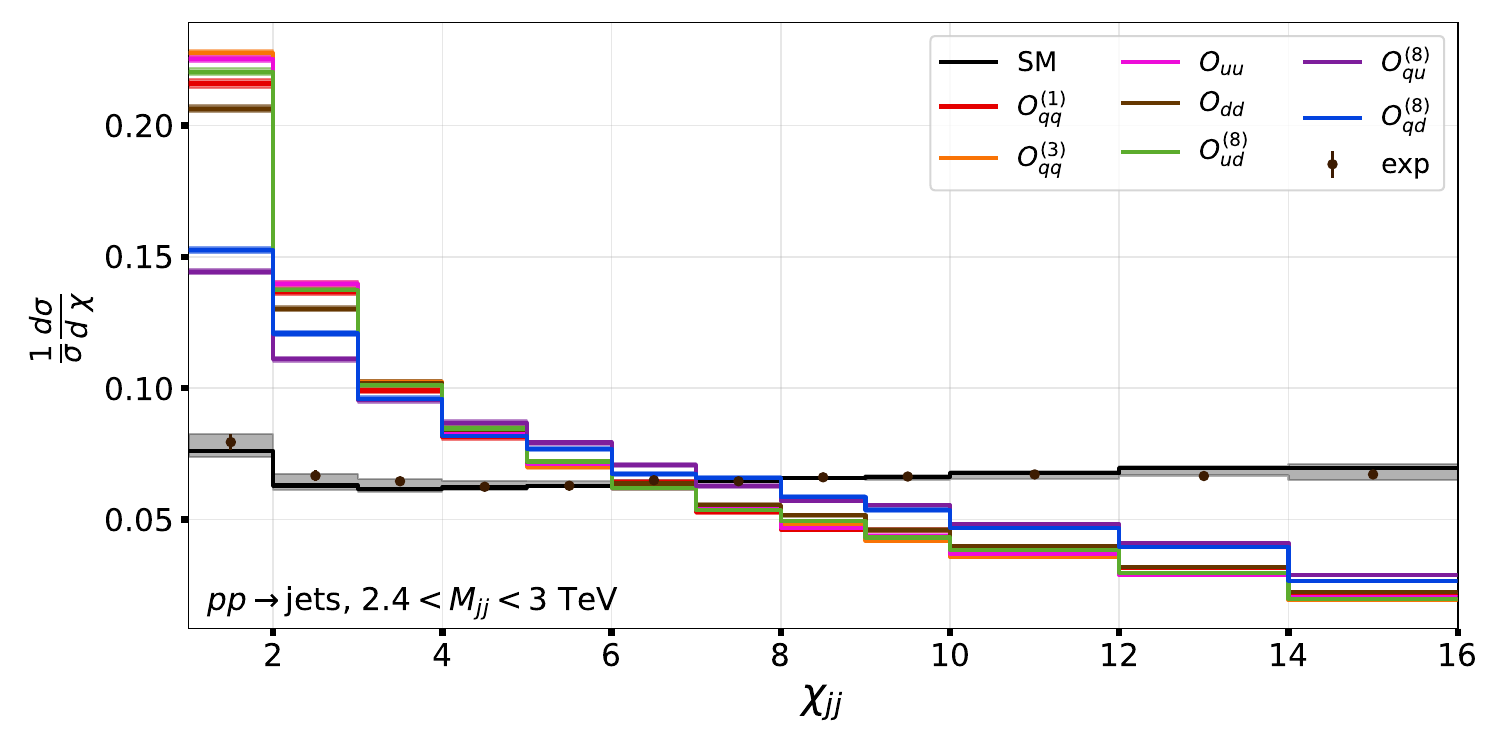}
   \includegraphics[width=.49\textwidth]{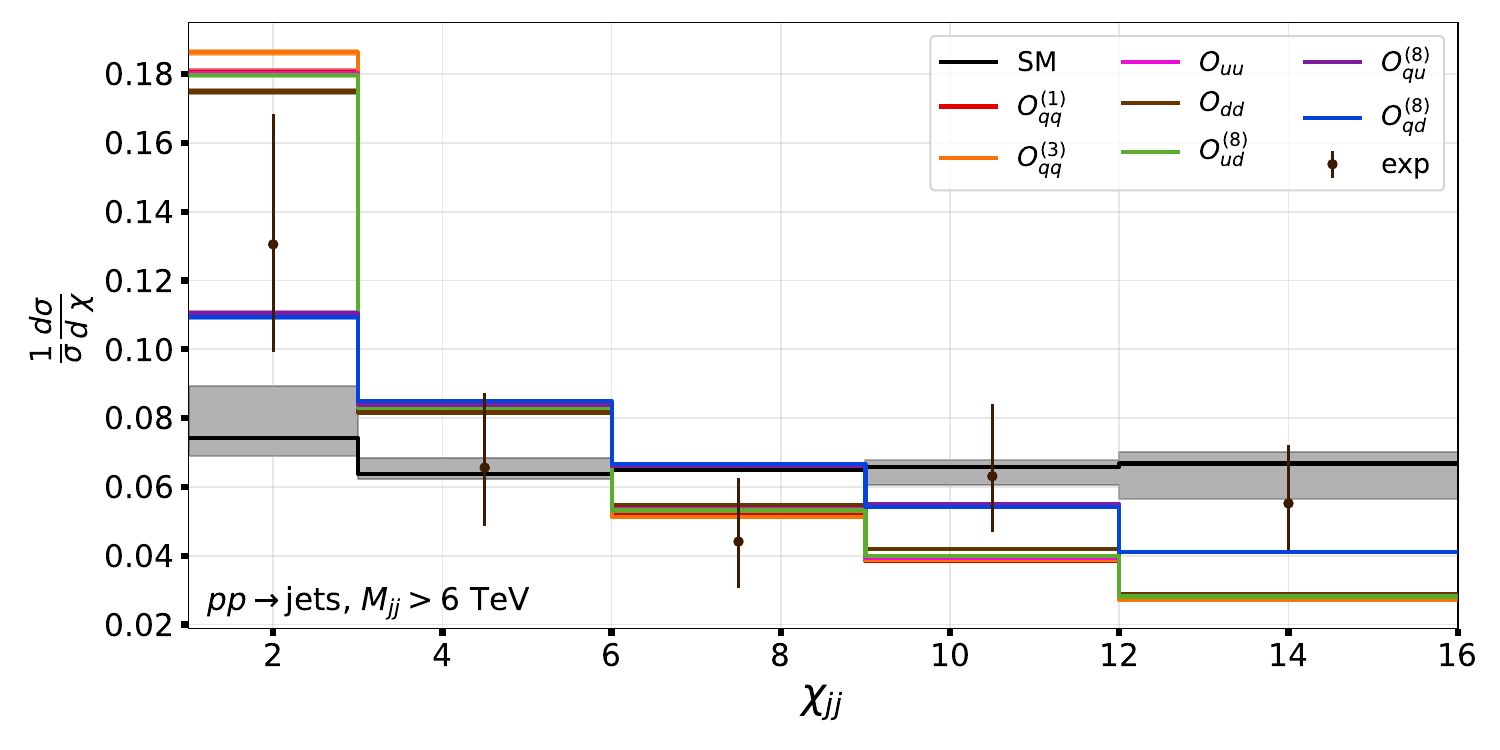}
\end{figure*}

\begin{figure*}
   \caption{\small{Differential distributions of the transverse momentum of the $b$-tagged jets in the $b+$jets region for multijet production, for the SM ({\it black}) and the contributing operators at LO, with all the coefficients set to $C_i/\Lambda^2 = 1$ TeV${}^{-2}$. The numerical uncertainties are represented with shaded bands, while the scale variations are shown in hatched bands for the SM only. The last bin includes the overflow}} \label{fig:bj_pTb}
   \includegraphics[width=.7\textwidth]{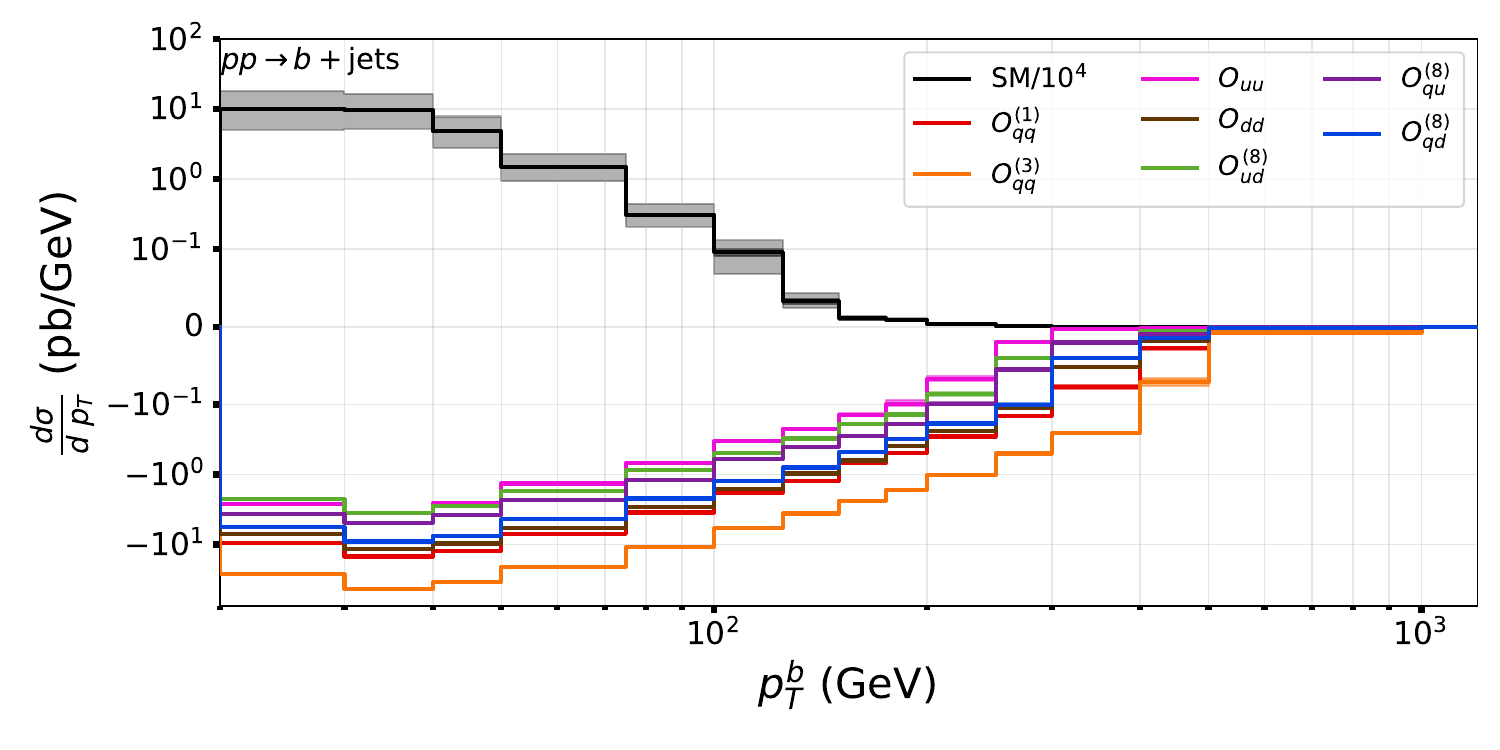}
\end{figure*}

\section{Multijet production}
These are among the most common processes at the LHC and the final states can easily reach the highest energies possible for that machine \cite{jets_1,jets_2,jets_3}. A study of the effects of the $O_G$ and some 4LQ operators in multijet processes can be found in \cite{Krauss_2017}.

We call $y_1,y_2$ the rapidities of the two leading jets in $p_T$ and define
\begin{align}
   \chi_{jj} &= e^{|y_1-y_2|}, \\
   y_\text{boost} &= \frac{|y_1+y_2|}{2}.
\end{align}
$\chi_{jj}$ is related to the scattering angle in the centre-of-mass (CoM) frame and the differential cross section as a function of it shows an almost flat trend for the SM, as the $t$-channel exchange of a gluon dominates and is independent of $|y_1-y_2|$. The 4LQ operators, on the other hand, produce a more isotropic angle distribution and show a peak of events for $\chi_{jj} \sim 1$.

A gluon cannot couple to quarks with different weak charges and same colour, so $O_{ud}^{(1)}$, $O_{qu}^{(1)}$ and $O_{qd}^{(1)}$ do not interfere with SM QCD and they are not considered for this process, as the QED term is usually subleading \cite{2j_SMEFT}. Analogously, the $O_{qq}^{(1)}$ interference does not contribute to subprocesses that involve both up- and down-like quarks. 

As suggested in the same reference, by employing \textsc{FeynArts} \cite{FeynArts} and \textsc{FormCalc} \cite{FormCalc} we saw that for each subprocess there are only two possible shapes for the interference differential $d\sigma^{1/\Lambda^2} / d \chi_{jj}$ cross sections at parton level: $O_{qq}^{(1)}$, $O_{qq}^{(3)}$, $O_{uu}$, $O_{dd}$ and $O_{ud}^{(8)}$ interferences with the SM generate the same trend, up to normalisations, while $O_{qu}^{(8)}$ produces the same as $O_{qd}^{(8)}$. The two distributions are shown in Fig. \ref{fig:dxsdChi} for the dominant channel $uu \rightarrow uu$, together with the SM one. Their analytical expressions can be found in Appendix \ref{app:dijet}. 

\subparagraph{Calculational details}
The phase space that we consider is defined as in the CMS analysis \cite{2j_2018}: we require $\chi_{jj}<16$, $|y_\text{boost}|<1.11$ and the invariant mass of the two leading jets has to satisfy $M_{jj}>2.4$ TeV. The CT18 PDF set \cite{ct18pdf} is used.

It is known that PS algorithms have problems with multijet events, as scales drop quite quickly and colour reconnection tends to spread radiation over large areas around the main interaction \cite{2j_NLO}. Because of this, a consistent fraction of events is lost when cuts are reapplied after PS, especially for the less-energetic SM jets. 
Even though the cross sections before the full matching and merging are not physical, it can be seen that their ratios before and after PS are similar for all the interference terms, which suggests that the events that are cut do not contain specific flavours in the final or initial states and that the kinematical differences induced by the PDF among the subprocesses do not affect the rejection rates.

The experimentally-measured and SM total cross sections in \cite{2j_2018} are not publicly available, so we assume for both of them the values we obtain from {\sc MadGraph} for the SM at LO, and use them to rescale the normalised CMS distributions. These numbers can be found in Table \ref{tab:all_xsects} for the first and last of the $M_{jj}$ regions in the experimental analysis: [2.4, 3] and [6, 13] TeV. The experimental and SM uncertainties from CMS are also rescaled by these factors when included in the $\chi^2$ to get limits. 

\subparagraph{Results}
The cross sections and uncertainties for the SM and each contributing operator are summarised in Table \ref{tab:all_xsects}, in the two $M_{jj}$ regions mentioned above. 
As it is often the case for four-fermion operators, their interference is negative for positive value of their coefficients. This can be understood from the fact that the integration of a heavy particle usually induces a negative coefficient for the corresponding four-fermion operator. For large values of the coefficients the cross section can therefore become negative, if only the interference is added. However, this issue is mitigated once constraints on the operator coefficients are taken into account, instead of the practical value of 1 TeV${}^{-2}$. Another possibility is to keep all the terms of the amplitude when squaring it, and not just the interference alone, but this introduces a partial contribution at the next order in the cutoff scale. As it can be seen in Table \ref{tab:all_xsects}, the same comments can be made for all the processes studied in this paper.

To get limits on the 4LQ Wilson coefficients, we consider the differential cross section $d\sigma / d\chi_{jj}$, computed in the [2.4, 3] and [6, 13] TeV $M_{jj}$ intervals; for $\chi_{jj}$, the binning we use is [1,2,3,4,5,6,7,8,9,10,12,14,16] in the first dijet mass region and [1,3,6,9,12,16] for the other. The $d\sigma / d\chi_{jj}$ LO SM distributions have been checked against the CMS predictions, that are normalised to 1, obtained through \textsc{NLOJET}++ \cite{nlojet}: ours exhibit lower tails, at higher values of $\chi_{jj}$.

The differential and normalised distributions for $\chi_{jj}$, matched to PS, are shown in Fig. \ref{fig:jj_chijj}, in the two $M_{jj}$ regions we consider, for the SM and the 4LQ operators that contribute. The two different shapes described above for $uu \rightarrow uu$ are still visible, one more peaked at $\chi_{jj} \sim 1$ for $O_{qq}^{(1)}$, $O_{qq}^{(3)}$, $O_{uu}$, $O_{dd}$ and $O_{ud}^{(8)}$, and one more flat and closer to the SM one for $O_{qu}^{(8)}$ and $O_{qd}^{(8)}$. This confirms that the PDF only favours different flavour mixes at different momentum fractions. It also proves that only two directions in the coefficient space can be probed by this process despite the number of bins, because the operators can only generate two distinct shapes for the differential cross section.

The experimental uncertainties lie between 1 and 8\% in the $2.4 < M_{jj} <3$ TeV region, and between 50 and 70\% in the $M_{jj} > 6$ TeV one. For the SM, they spread up to 12\% and 30\% in the same invariant-mass intervals.

The largest correction to the SM, when all the Wilson coefficients are equal, comes from $O_{qq}^{(3)}$ as it is the only operator that contributes to all the subprocesses. Scale variations for this sample are around 35\% in the [2.4, 3] TeV region and 60\% in the [6, 13] TeV one. Numerical uncertainties are of order 1\% in both.

\subparagraph{Multijets with $b$- and $c$-tagging}
By applying the $b$- and $c$-tagging algorithms described in Sec. \ref{sec:framework}, we check the contributions of the 4LQ operators to $b+$jets, $c+$jets and $bb+$jets productions, where respectively at least one $b$-jet, at least one $c$-jet and at least two $b$-jets are identified. The request for a $b$-jet to be present in the final state, for example, increases the sensitivity to operators that feature down-like quarks, and analogously holds for $c$-jets and up-like ones. The operators that contribute at linear level are the same as above.

Jets are included in the analysis if they show $p_T^j>20$ GeV and $|y_j|<2.5$. Due to the operational window of the tagging algorithms, jets cannot be $b$-tagged if they show $p_T > 3$ TeV, or $c$-tagged with $p_T > 250$ GeV.

For the experimental data, we assume the same differential distributions and uncertainties as the LO SM ones that we obtain through {\sc MadGraph}.

The total cross sections are summarised in Table \ref{tab:tag_xsects} in the Appendix. It is possible to observe that the requirement of a $b$-jet in the final state, for example, does not rule out the operators that only feature up-like quarks, like $O_{uu}$, due to the mistagging of $c$- and lighter-jets.

The variable we consider for extracting limits, in this case, is the transverse momentum $p_T^b$ of the $b$-tagged jets in the $b+$jets region. Its differential distributions can be observed in Fig. \ref{fig:bj_pTb}. The SM relative scale variations that we obtain are larger than 60\%, and in the first bins they increase above 100\%. As in the other multijets regions, the primary correction to the SM comes from $O_{qq}^{(3)}$, if all the coefficients are set to 1 TeV${}^{-2}$. For this operator, the scale variations lie between 50\% and 80\% in all bins but the first ones, where they get over 100\%.

Due to the relatively low scale in $p_T$ that the events can probe, the NP contribution is much smaller, compared to the SM one, than in the generic multijet production discussed above: therefore, constraints are expected to be rather weak.

\begin{figure*}
   \caption{\small{Differential ({\it top}) and normalised ({\it bottom}) distributions of the azimuthal distance between the two leading jets in $\ell^+ \ell^- +$jets production, for the SM ({\it black}) and the interference of the ten operators included in this analysis at LO, with all the coefficients set to $C_i/\Lambda^2 = 1$ TeV${}^{-2}$. The numerical uncertainties are represented with shaded bands in both plots, while the scale variations are shown in hatched bands in both plots for the SM, and only on top for the 4LQ operators. The experimental measurements are also included}} \label{fig:lljj_Dphijj}
   \includegraphics[width=.7\textwidth]{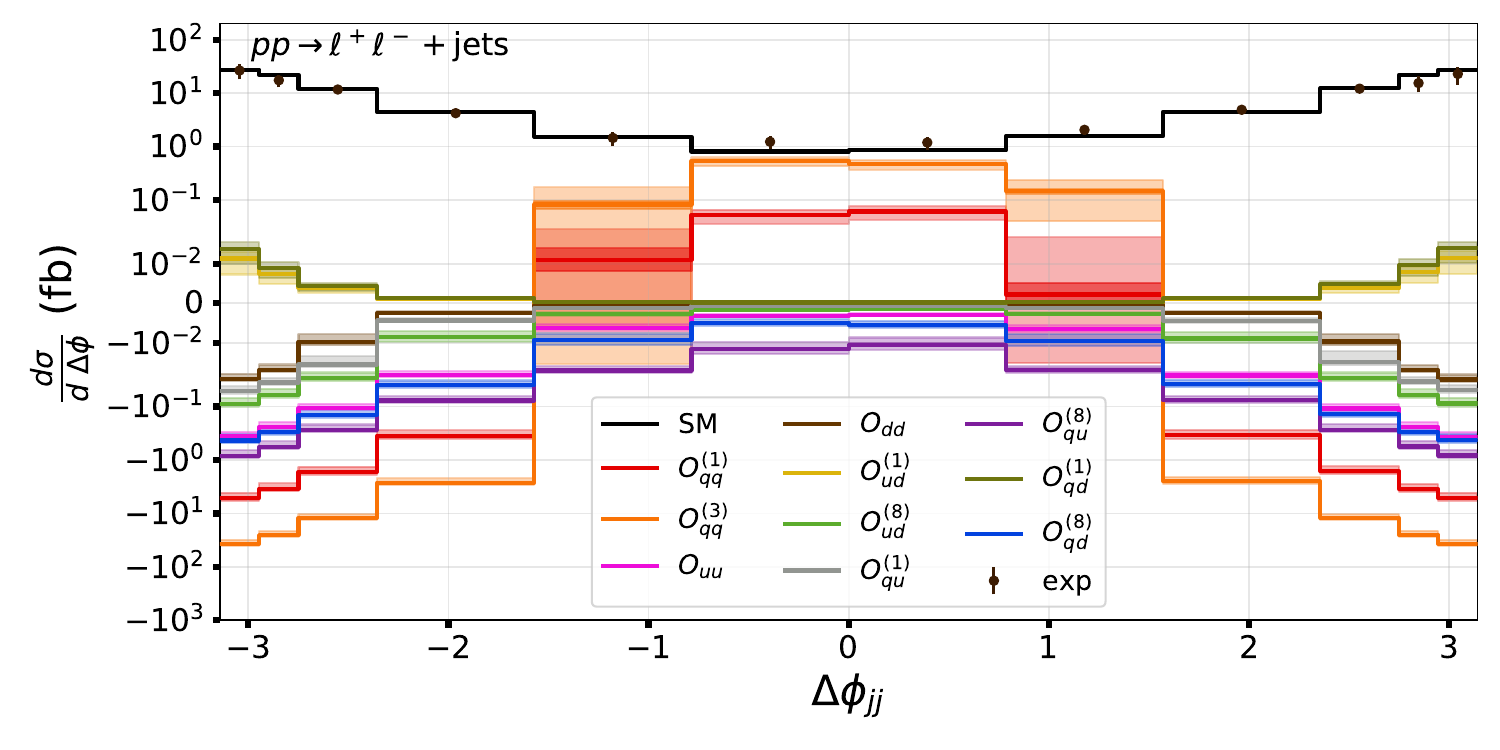}
   \includegraphics[width=.7\textwidth]{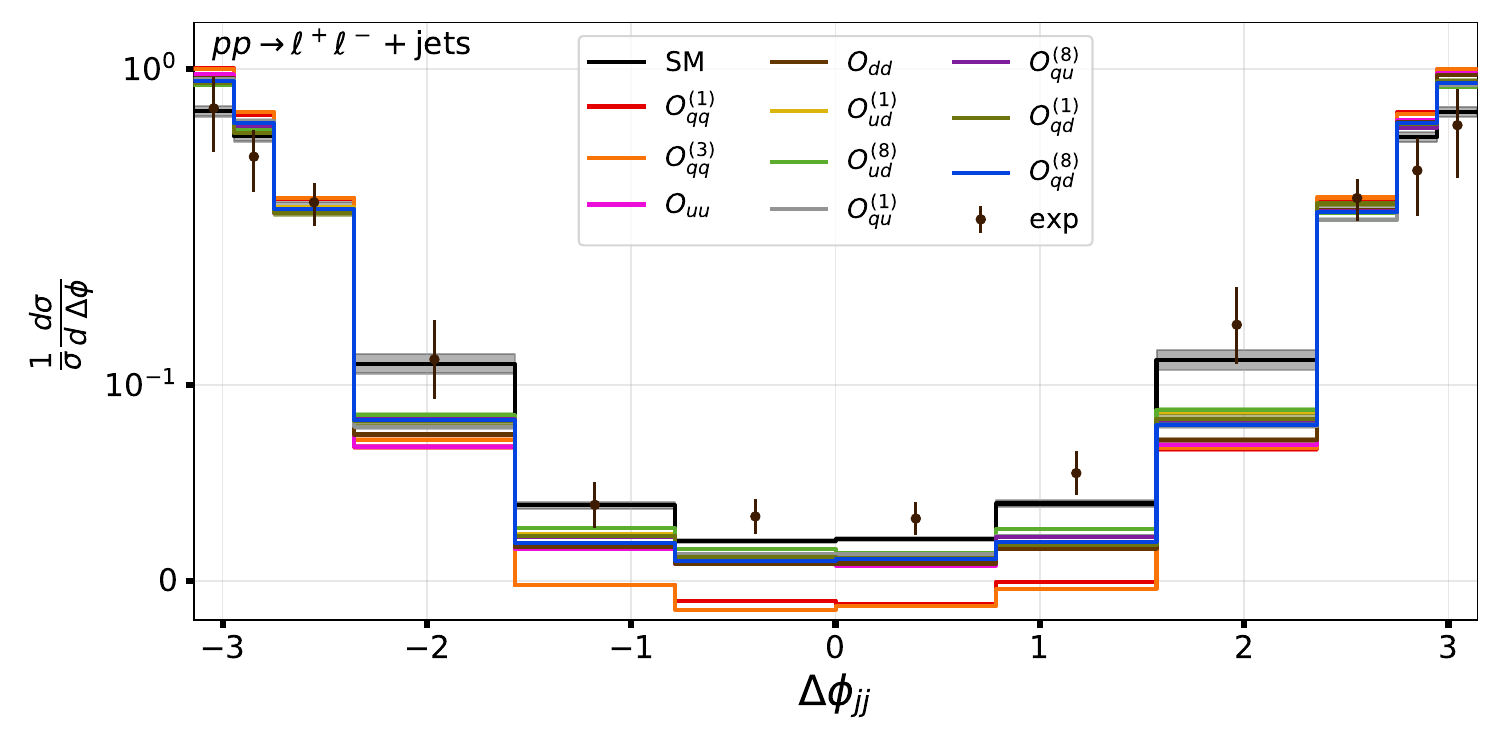}
\end{figure*}

\section{{\it Z}+jets production}
The relevance of processes where jets are produced together with EW bosons has increased in the past years, not only because they represent a background to the Higgs boson production via VBF but also for new physics searches \cite{bsm_zjj_1,bsm_zjj_2,bsm_zjj_3}. We target the interference among the SM diagrams and the four-fermion interaction ones, where one fermionic line emits a $Z$ boson that decays into a pair of leptons ($Z \rightarrow \ell^+ \ell^-$). All the ten operators included in our analysis contribute to this process at $\mathcal{O}(1/\Lambda^2)$ order, as it can be seen in Table \ref{tab:ops_def}.

\subparagraph{Calculational details}
We attempt to recast the ATLAS analysis \cite{Atlas:2020zjj}. Two leptons with same flavour and opposite electric charge have to be present, with $p_T^\ell>25$ GeV, rapidity satisfying $|y_\ell|<2.5$, invariant mass $M_{\ell\ell}$ in the interval [81, 101] GeV and $p_T^{\ell\ell}>20$ GeV.

The reconstructed jets are considered if they show $p_T^j>25$ GeV, $|y_j|<4.4$ and minimum $\Delta R_{\ell j}$ of 0.4 with the leptons; at least two jets need to be present. The two leading jets in $p_T$ have to show transverse momenta above 85 and 80 GeV respectively, $M_{jj}>1$ TeV and $|\Delta y_{jj}|>2$. Furthermore, no other jet must be present in the rapidity gap among them. Finally, to ask for the $Z$ boson to be centrally produced with respect to the jets, we impose $\xi_Z<0.5$, with the last quantity defined as
\begin{equation}
   \xi_Z = \frac{|y_{\ell\ell}-\frac{1}{2}(y_{j1}+y_{j2})|}{|\Delta y_{jj}|}.
\end{equation}
$y_{\ell\ell}$, $y_{j1}$ and $y_{j2}$ stand for the rapidities of the dilepton system and of the two leading jets. In particular, the last cuts are requested in the ATLAS analysis to enhance the EW VBF contribution to the process. While other cuts may be more efficient to target 4LQ operators, we limit ourselves to the experimental setup as a first step, keeping the investigation of different phase-space areas for future work.

NNPDF3.0 is used as PDF, with $\alpha_S (M_Z)=0.118$ \cite{nnpdf30}. It is known that predictions from different generators do not agree for the VBF processes, even for the SM, and that results heavily depend on the shower choice \cite{wwjj_atlas,qcd_rad_vbf,ps_vbf,cw_paper}. In the reference analysis, the SM from {\sc Herwig7+Vbfnlo} seems to be in closest agreement with the total cross section measured experimentally, so these are the distributions that are assumed for the SM. Anyway, we would like to stress that no MC generator considered in that study can match the measured differential distributions everywhere, and that choosing the SM predictions that are closest to the data might fit possible room for new physics away.

\subparagraph{Results}
The inclusive cross sections and uncertainties for the SM and the interference of each operator are summarised in Table \ref{tab:all_xsects}.

The variables we investigate are the azimuthal distance between the leading jets $\Delta\phi_{jj}$, their invariant mass $M_{jj}$ and absolute rapidity difference $|\Delta y_{jj}|$, and the transverse momentum of the dilepton system $p_T^{\ell\ell}$. 

The differential distributions and their shapes for the first one are shown in Fig. \ref{fig:lljj_Dphijj}, for the SM and all the operators included in this study, together with the experimental data. The bins we use are [0, $\pi$/4, $\pi$/2, 3$\pi$/4, 7$\pi$/8, 15$\pi$/16, $\pi$] and their symmetric around 0. 

The uncertainties on the SM generation presented in the ATLAS analysis are close to 9\% in every bin. In the experimental measurements, the largest source of uncertainty comes from the comparison among different MC generators, for the reasons stated above, and it reaches a 30\% impact alone in the external bins.

It can be seen that $O_{qq}^{(3)}$ produces the largest deviation from the SM, if all coefficients are set to 1 TeV${}^{-2}$, and that most of the interference corrections are negative, but a sign flip occurs in the central bins for $O_{qq}^{(1)}$ and $O_{qq}^{(3)}$. Relative scale variations in the $O_{qq}^{(3)}$ distribution are of order 25\% in all bins, except the two ones where the sign change happens, which show uncertainties above 100\%. The same two bins show larger numerical uncertainties ($\sim$ 20\%) compared to the others ($\sim$ 1\%).

\begin{figure*}
   \caption{\small{LO differential ({\it top}) and normalised ({\it bottom}) distributions of the azimuthal distance among the two leading jets in $\ell^\pm \nu+$jets production, for the SM ({\it black}) and the operators that contribute, with all the coefficients set to $C_i/\Lambda^2 = 1$ TeV${}^{-2}$. The numerical uncertainties are represented with shaded bands in both plots, while the scale variations are shown in hatched bands in both plots for the SM, and only on top for the 4LQ operators}} \label{fig:lvjj_Dphijj}
   \includegraphics[width=.7\textwidth]{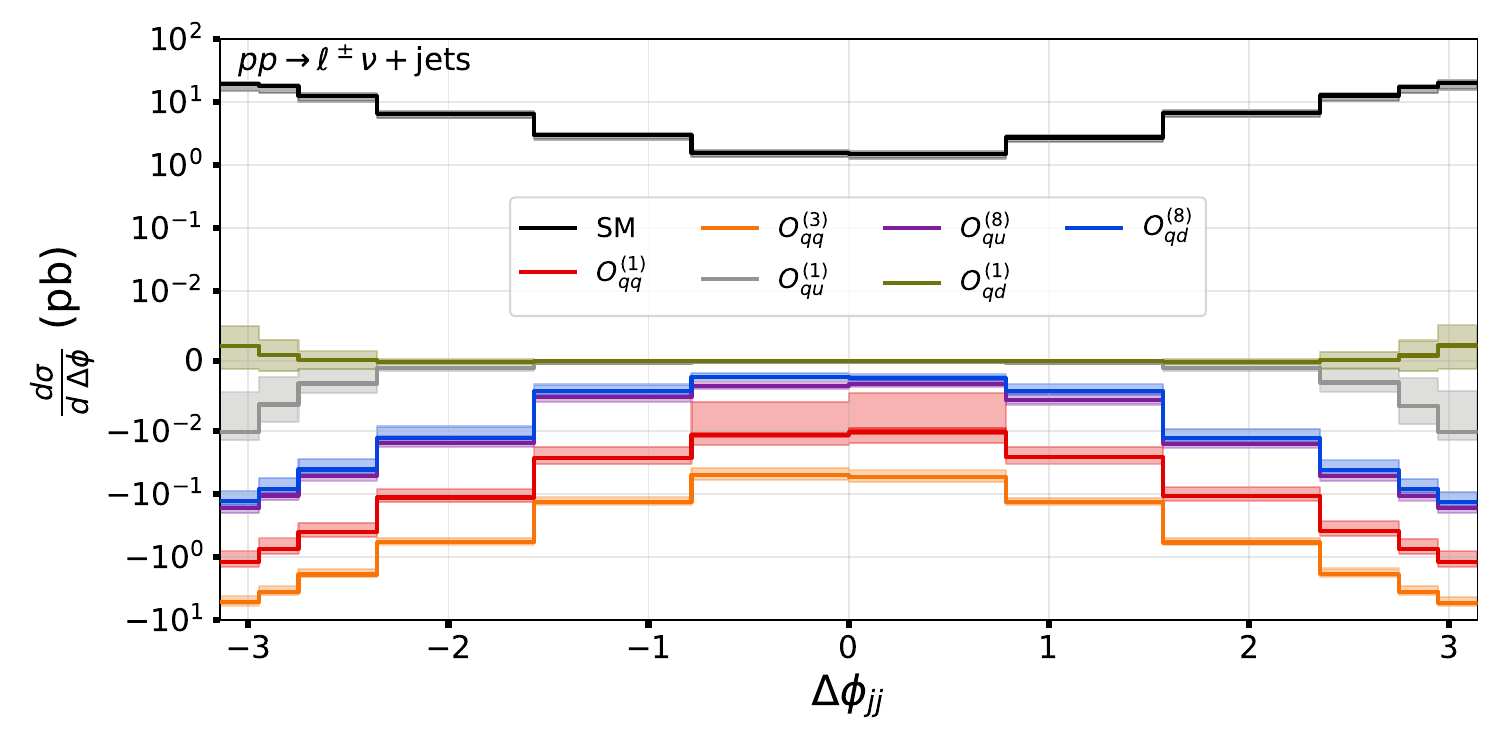}
   \includegraphics[width=.7\textwidth]{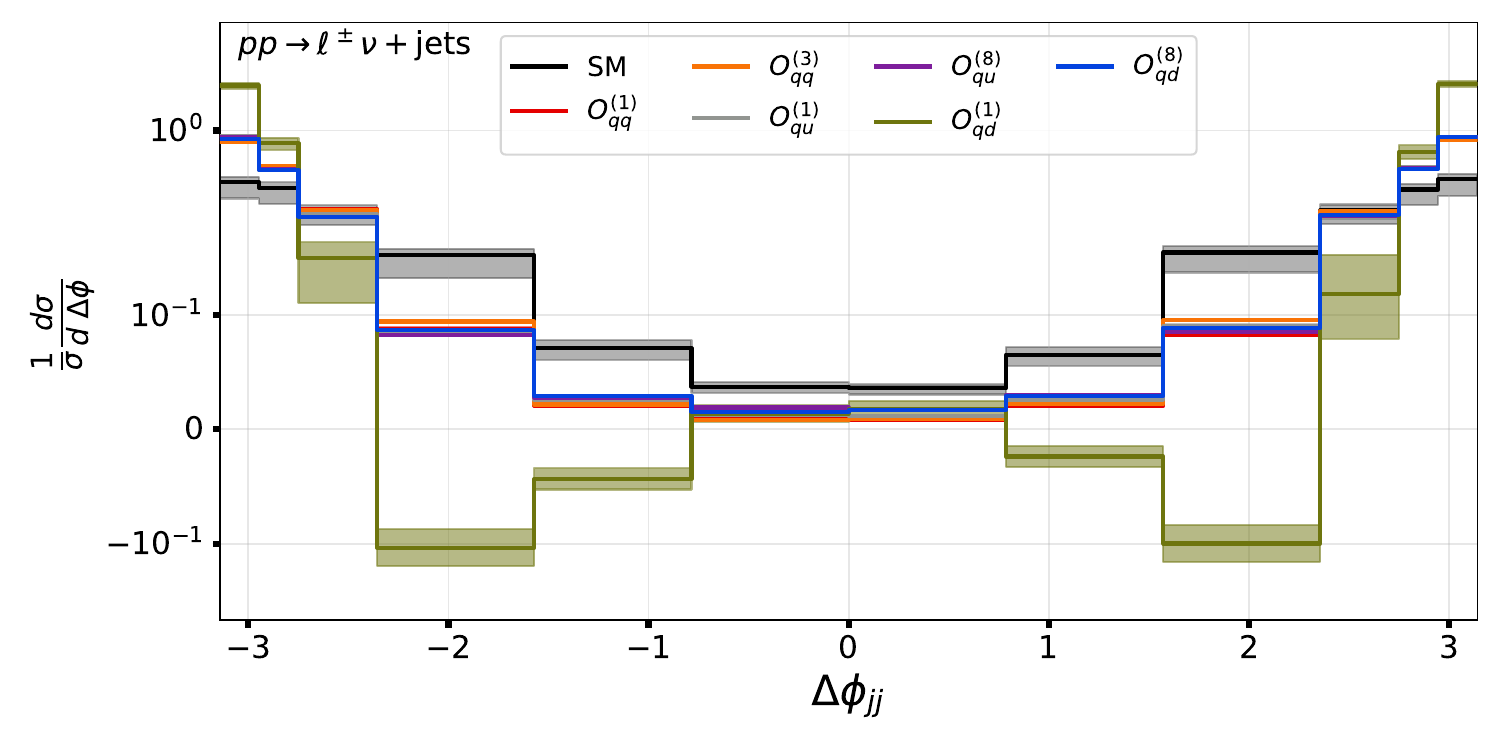}
\end{figure*}

\section{{\it W}+jets production}
In general, the production of a $W$ boson in association with jets is studied in the VBF regime to investigate the SM, model the parton radiation, and NP studies, like the generation of anomalous trilinear gauge couplings \cite{bsm_wjj_1,bsm_wjj_2,bsm_wjj_3}. As in the previous section, our diagrams of interest include a four-fermion interaction with the emission of a leptonically-decaying $W$ boson from one of the quark lines ($W^\pm \rightarrow \ell^\pm \nu$). Due to the left-handed nature of the $W$-boson interaction, $O_{uu}$, $O_{dd}$, $O_{ud}^{(1)}$ and $O_{ud}^{(8)}$ do not contribute, as it is summarised in Table \ref{tab:ops_def}.

\subparagraph{Calculational details}
The cuts we apply are the ones imposed in the CMS analysis \cite{cms:2020wjj}: events need to contain exactly one lepton, with $p_T^\ell>27$ GeV and $|y_\ell|<2.4$, and at least two jets, with the two leading in $p_T$ satisfying $M_{jj}>200$ GeV and with transverse momenta of at least 50 and 30 GeV respectively. Other jets are considered if they show $p_T>25$ GeV and $|y|<5$. Both the missing transverse energy $p_T^\text{miss}$ and the transverse mass $M_T^{\ell\nu}$ of the lepton-neutrino system have to exceed 40 GeV. NNPDF3.0 is used as PDF in the generation.

The longitudinal component of the neutrino momentum is estimated from $p_T^\text{miss}$ by assuming that the $W$ boson is on-shell. Among the possible solutions that this condition yields, the one with the lower absolute value is considered \cite{Rahaman:2020}. If no real roots can be found, we obtain one by discarding the imaginary part.

As a Boosted Decision Tree (BDT) is trained in the experimental analysis to separate the EW contribution from the rest, we generate the SM at LO with {\sc MadGraph} and assume that the experimental data will agree with the SM distributions. For each bin, we introduce an experimental uncertainty equal to 10\%, similarly to the central region in the SM $\Delta \phi_{jj}$ distribution for $Z+$jets in the previous section: the $W+$jets cross section is larger, so we expect smaller relative error bars.

\subparagraph{Results}
The total cross sections for the SM and the interference of each operator are listed in Table \ref{tab:all_xsects}.

We check the following observables: $\Delta\phi_{jj}$, $M_{jj}$ and $|\Delta y_{jj}|$ among the two leading jets, the transverse momenta $p_T^{j1}$, $p_T^{j2}$ and $p_T^\ell$, $M_T^{\ell\nu}$ for the lepton-neutrino system, the angular distances $\Delta R_{W, j1}$ and $\Delta R_{\ell, j1}$ among the leading jet and the $W$ boson or lepton, the triple product $\frac{(\vec{j}_1\times \vec{j}_2)\cdot \vec{\ell}}{|\vec{j}_1\times \vec{j}_2||\vec{\ell}|}$, and the azimuthal angle $\phi_W$ between the plane containing the $W$ boson and the beam axis, and the plane of the lepton and the neutrino, in lab frame \cite{Azatov_2019}. 

The differential distributions and the shapes for $\Delta\phi_{jj}$ are shown in Fig. \ref{fig:lvjj_Dphijj}, for the LO SM and the operators that contribute to this process. The bins we use are [0, $\pi$/4, $\pi$/2, 3$\pi$/4, 7$\pi$/8, 15$\pi$/16, $\pi$] and their symmetric around 0, as in the $Z+$jets case. 
In our SM distribution, numerical uncertainties are of order 1\% and scale ones lie between 25 and 30\%. The largest correction comes from $O_{qq}^{(3)}$, if all coefficients are equal. The relative scale variations in this sample are between 25 and 50\%. $O_{qd}^{(1)}$ produces a different shape from the other operators, but its contribution to the cross section is negligible.

Also $\Delta R_{W, j1}$ is a valid observable for this process, as it is sensitive to the jet chiralities. It yields slightly worse individual bounds on the operator coefficients than $\Delta \phi_{jj}$, because of the larger uncertainties over the SM predictions and the smearing induced by the $W$-boson reconstruction.

\begin{figure*}
   \caption{\small{Differential ({\it top}) and normalised ({\it bottom}) distributions of the photon transverse momentum in $\gamma+$jets production, for the LO SM ({\it black}) and the 4LQ operators, with all the coefficients set to $C_i/\Lambda^2 = 1$ TeV${}^{-2}$. The numerical uncertainties are represented with shaded bands in both plots, while the scale variations are shown in hatched bands in both plots for the SM, and only on top for the 4LQ operators. The experimental measurements are also shown. The last bin contains the overflow}} \label{fig:ajj_pTa}
   \includegraphics[width=.7\textwidth]{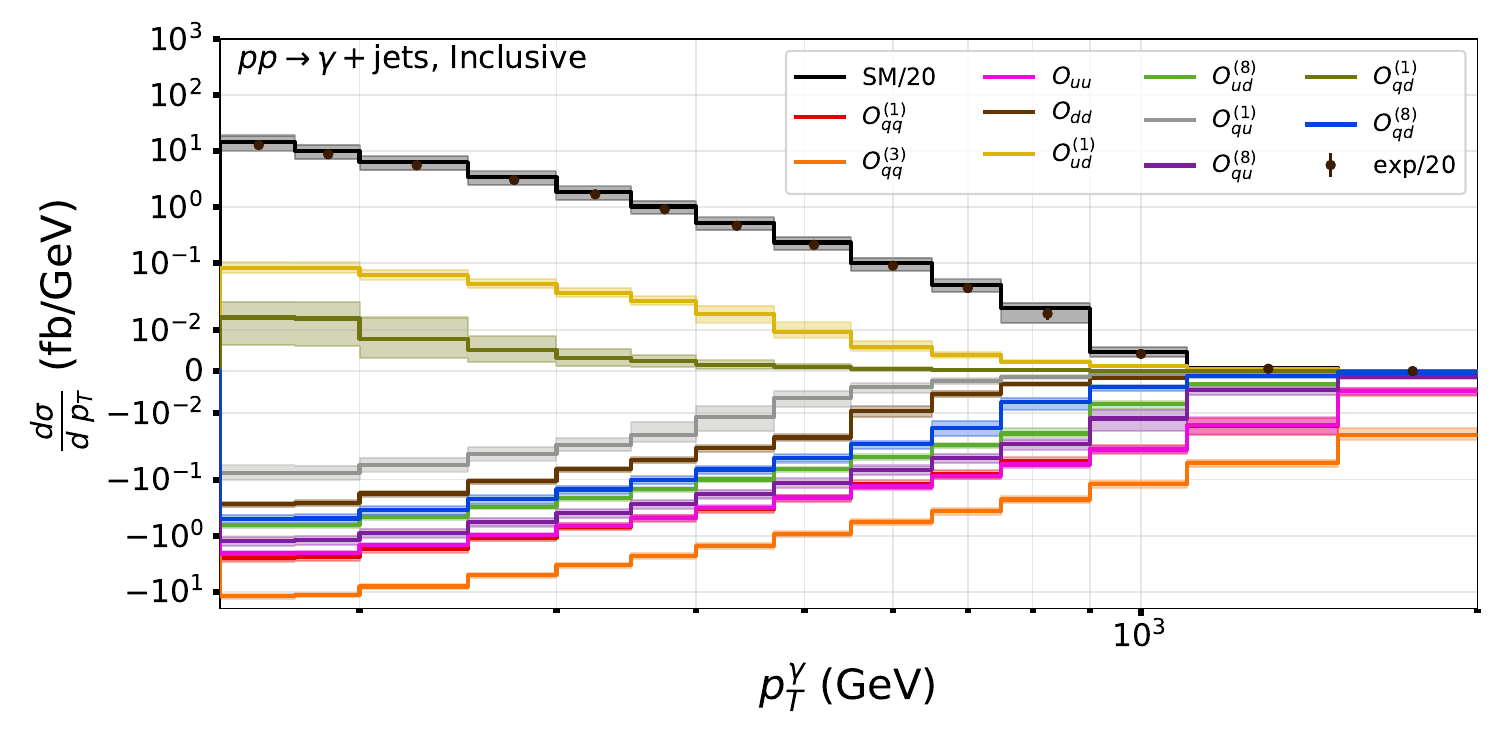}
   \includegraphics[width=.7\textwidth]{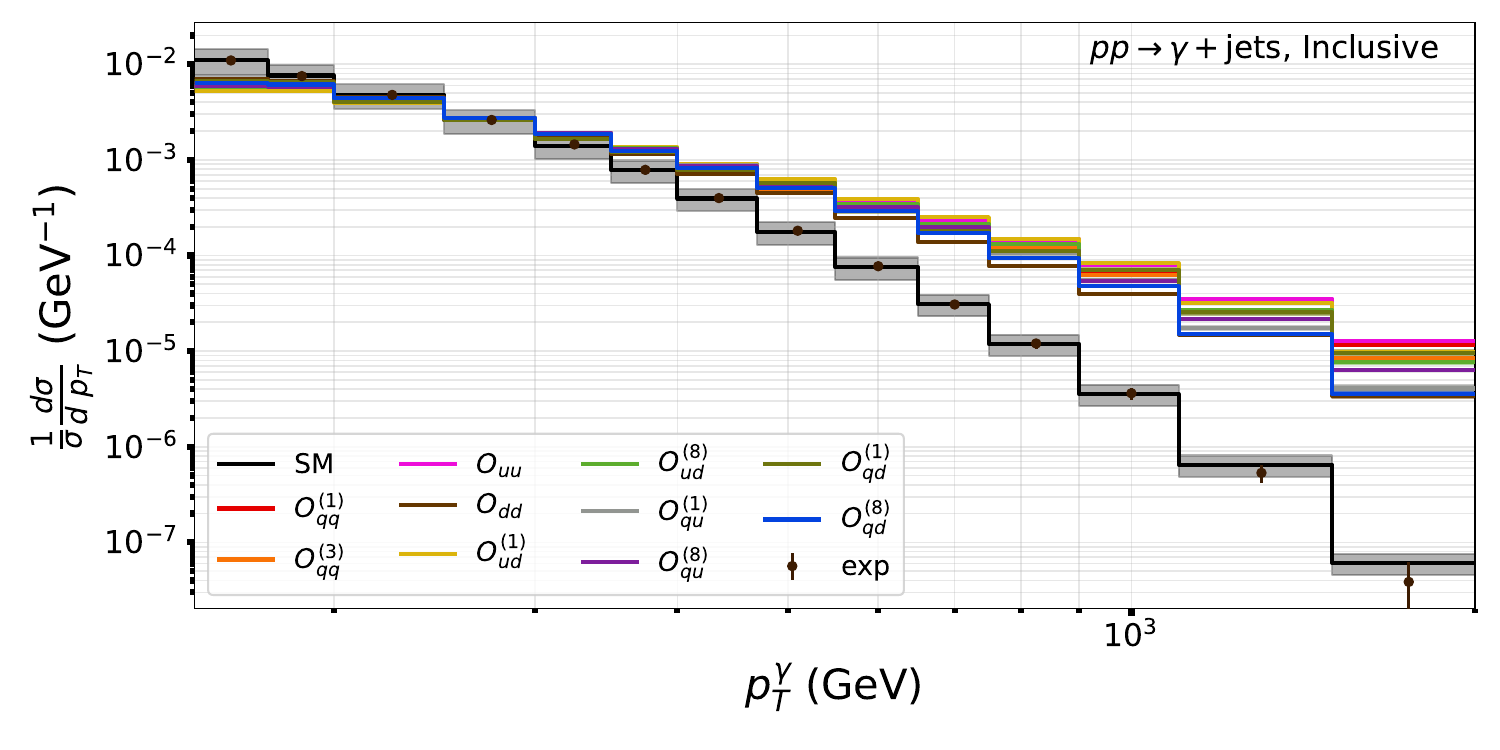}
\end{figure*}

\section{\label{sec:ajj}Photon+jets production}
The production of a photon in association with jets is studied at the LHC to test perturbative QCD. The photon can originate by the hard interaction (direct process) or by the fragmentation of a high-$p_T$ parton (fragmentation process); the latter is usually distinguished from a hadron-in-jets decay into photons thanks to isolation requirements.

As in the previous processes, we are interested in the interference among the SM and four-fermion interactions induced by 4LQ operators, where one fermionic line emits a photon. All the operators we are considering contribute to this process. In principle, due to the different electric charges, the $\mathcal{O}(1/\Lambda^2)$ term should be enhanced when the photon is emitted by an up-like quark rather than a down-like one.

\subparagraph{Calculational details}
Our phase space is defined as in the ATLAS analysis \cite{Atlas:2020ajj}. At least one photon with $p_T^\gamma>150$ GeV and $|y_\gamma|<2.37$ is requested, together with at least two jets with a minimum $p_T^j$ of 100 GeV and a maximum $|y_j|$ of 2.5. An angular separation $\Delta R_{\gamma j}>0.8$ is required between the photon and leading jets. For the other jets, only the ones with $p_T>20$ GeV and $|y|<2.5$ are considered.

Together with an inclusive region delimited by these cuts, a direct-enriched ($p_T^\gamma > p_T^{j1}$) and a fragmentation-enriched ($p_T^\gamma < p_T^{j2}$) ones are also defined, in order to investigate the two topologies described above.

NNPDF3.0 is used for PDF. The SM predictions that we include are the {\sc Sherpa} ones from the reference analysis, and we compare them against the experimentally-measured distributions in the same paper \cite{sherpa}.

\subparagraph{Results}
The inclusive cross sections for the SM and the interference terms can be found in Table \ref{tab:all_xsects}; the cross sections in the fragmented- and direct-enriched regions are listed in Appendix \ref{app:ajj}. $O_{ud}^{(1)}$, $O_{qu}^{(1)}$ and $O_{qd}^{(1)}$ do not interfere with the $t$-channel gluon exchange diagram, that dominates the SM contribution, so their cross sections are smaller than the others \cite{smeft_global_mfv}. 

We comput predictions, in all three phase-space regions, for the following observables: the transverse momenta of the photon and jets, $p_T^\gamma$ and $p_T^j$, the absolute rapidity of the jets $|y_j|$, the rapidity and azimuthal distances of the leading jets among them and the photon, $|\Delta y_{jj}|$,  $|\Delta \phi_{jj}|$, $|\Delta y_{\gamma j}|$, $|\Delta \phi_{\gamma j}|$, and the invariant masses $M_{jj}$ and $M_{\gamma jj}$ of the dijet and photon-jet-jet systems.

The differential and normalised contributions of the SM and the 4LQ interference terms to $p_T^\gamma$ are shown in Fig. \ref{fig:ajj_pTa}, together with the experimental measurements. The binning we employ is [150, 175, 200, 250, 300, 350, 400, 470, 550, 650, 750, 900, 1100, 1500, 2000] GeV, with the last one containing the overflow. The uncertainties on the SM distribution lie between 50 and 60\% in all bins, while the experimental ones are of order 10\% in all bins but the last two, where they increase. The largest deviation from the SM rises, as in all the other processes, from $O_{qq}^{(3)}$: in the histogram we obtain, numerical uncertainties are of order 1\% in every bin, while scale ones vary between 15 and 30\%. A difference in shape can be observed between the SM and 4LQ contributions, as the latter show higher tails at larger $p_T$ values. However, here again all the operators share in good approximation the same shape.

In previous studies \cite{smeft_global_mfv, Atlas:2024ajj}, $M_{jj}$ is suggested as an observable to investigate this process, since it is sensitive to the hard-interaction dynamics. In our predictions, though, we find that the interference effect over the SM one is larger for $p_T^\gamma$ by at least a factor $\sim 4$ in the tail bins, leading to more stringent bounds. For this reason, this is the variable we adopt in our analysis.

\begin{figure*}
   \caption{\small{Individual limits, at 95\% CL at interference level, on the 4LQ operators from the processes and variables included in this study. The bounds at square level are shown for $C_{qq}^{(3)}$ only, in dashed. The combined constraints, in black, come from the inclusion of all the processes included in this study. For multijet production, the same observable is considered in two dijet invariant-mass regions separately, and the experimental and SM normalised distributions from CMS are rescaled by the total LO SM cross section we computed. In the $W+$jets case, the interference terms are compared against the LO SM generated by us. The numerical values for these limits are listed in Table \ref{tab:ind_limits}}} \label{fig:ind_limits}
   \includegraphics[width=.8\textwidth]{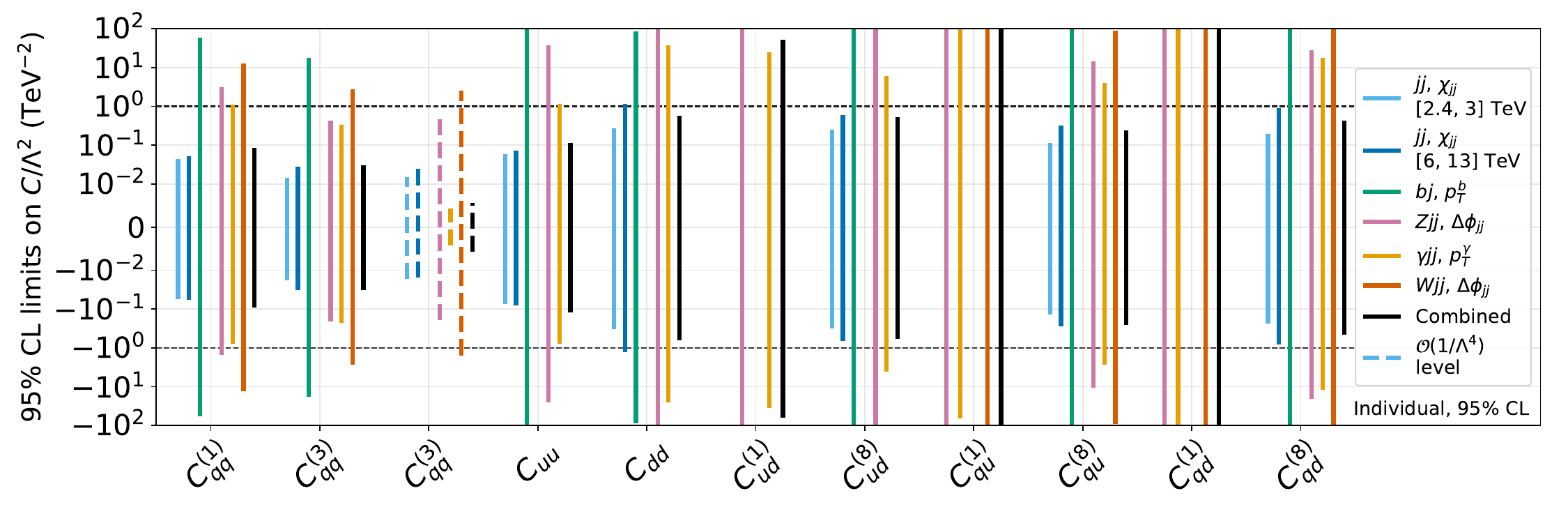}
\end{figure*}

\begin{table*}
\caption{\small{List of experimental measurements used to obtain bounds on the 4LQ coefficients, for the processes studied in this chapter. For $W+$jets and $b+$jets, we assume that the measurements follow the LO SM distribution. For multijets, we use the total LO SM cross section to multiply the normalised experimental distributions presented in the reference. The two $N_\text{data}$ values for multijets refer to the low- and high-$M_{jj}$ regions that we consider, respectively}} \label{tab:4lq_dataset}
\resizebox{.7\textwidth}{!}{
\begin{tabular}{cc|cccc}
   \hline
   Proc. & Observable & $\sqrt{s}$, $\mathcal{L}$ & Final state & $N_\text{data}$ & Ref. \\ \hline
   Multijets & $d^2 \sigma / (d\chi_{jj} \hspace{1mm} d M_{jj})$ & 13 TeV, 35.9 fb${}^{-1}$ & jets & 12, 5 & \cite{2j_2018} \\
   $b+$jets & $d \sigma /d p_T^b$ & \multicolumn{4}{c}{Exp. data taken as LO SM} \\
   $Z+$jets & $d\sigma / d\Delta \phi_{jj}$ & 13 TeV, 139 fb${}^{-1}$ & $\ell^+ \ell^- +$jets, $\ell=e,\mu$ & 12 & \cite{Atlas:2020zjj} \\
   $W+$jets & $d\sigma / d\Delta \phi_{jj}$ & \multicolumn{4}{c}{Exp. data taken as LO SM} \\
   $\gamma +$jets & $d\sigma /d p_T^\gamma$ & 13 TeV, 36.1 fb${}^{-1}$ & $\gamma +$jets & 14 & \cite{Atlas:2020ajj} \\ \hline
\end{tabular}
}
\end{table*}

\begin{table*}
\caption{\small{Best constrained directions, in the coefficient space, by the processes and main observables included in this study. The last column is obtained from the sum of all the other $\chi^2$ polynomials in the table. The eigenvalues $\lambda$ for each of them, with $1\sigma$ uncertainties, are also shown. The symbol $\times$ is used to mark the operators that do not contribute to a certain process, while zeros replace entries that are smaller than $10^{-2}$, in absolute value. Details on the calculations can be found in Sec. \ref{sec:framework}}} \label{tab:constr_dirs}
\resizebox{.8\textwidth}{!}{
\begin{tabular}{c|ccccccc}
\hline
\multicolumn{8}{c}{Best-constrained directions in coefficient space} \\ 
 & \multicolumn{2}{c}{Multijets, $\chi_{jj}$} & $b+$jets & $\ell^+ \ell^- +$jets & $\ell^\pm \nu +$jets & $\gamma+$jets & Combined \\
 & $2.4<M_{jj}<3$ TeV & $M_{jj}>6$ TeV & $p_T^b$ & $\Delta\phi_{jj}$ & $\Delta\phi_{jj}$ & $p_T^\gamma$ & \\ \hline
$C_{qq}^{(1)}$ & 0.41 & 0.52 & 0.39 & 0.25 & 0.36 & 0.38 & 0.43 \\
$C_{qq}^{(3)}$ & 0.83 & 0.74 & 0.83 & 0.97 & 0.93 & 0.84 & 0.82 \\
$C_{uu}$ & 0.32 & 0.40 & 0.02 & 0.02 & $\times$ & 0.37 & 0.33 \\
$C_{dd}$ & 0.07 & 0.03 & 0.28 & 0 & $\times$ & 0.01 & 0.07 \\
$C_{ud}^{(1)}$ & $\times$ & $\times$ & $\times$ & 0 & $\times$ & -0.01 & 0 \\
$C_{ud}^{(8)}$ & 0.08 & 0.05 & 0.10 & 0 & $\times$ & 0.07 & 0.07 \\
$C_{qu}^{(1)}$ & $\times$ & $\times$ & $\times$ & 0 & 0 & 0 & 0 \\
$C_{qu}^{(8)}$ & 0.17 & 0.10 & 0.12 & 0.04 & 0.05 & 0.11 & 0.16 \\
$C_{qd}^{(1)}$ & $\times$ & $\times$ & $\times$ & 0 & 0 & 0 & 0 \\
$C_{qd}^{(8)}$ & 0.10 & 0.04 & 0.22 & 0.02 & 0.04 & 0.03 & 0.09 \\ \hline
$\lambda$ & 1.1$\cdot 10^5$ & 2.1$\cdot 10^4$ & 1.2$\cdot 10^{-1}$ & 1.6$\cdot 10^2$ & 3.1 & 4.4$\cdot 10^2$ & 1.3$\cdot 10^5$ \\
 & $\pm 50\%$ & $\pm 57\%$ & $ \pm 67\%$ & $\pm 40\%$ & $\pm 52\%$ & $\pm 50\%$ & $\pm 54\%$ \\
\hline
\end{tabular}
}
\end{table*}

\begin{figure*}
   \caption{\small{Individual contours in the $C_{qq}^{(1)}$ {\it vs} $C_{qq}^{(3)}$ ({\it left}) and $C_{qu}^{(8)}$ {\it vs} $C_{qq}^{(3)}$ ({\it right}) planes. The contours from $b+$jets and $W+$jets are not shown as they exceed the limits on the axes. Note that the ranges are different on the two axes}} \label{fig:ind_contours}
   \includegraphics[width=.49\textwidth]{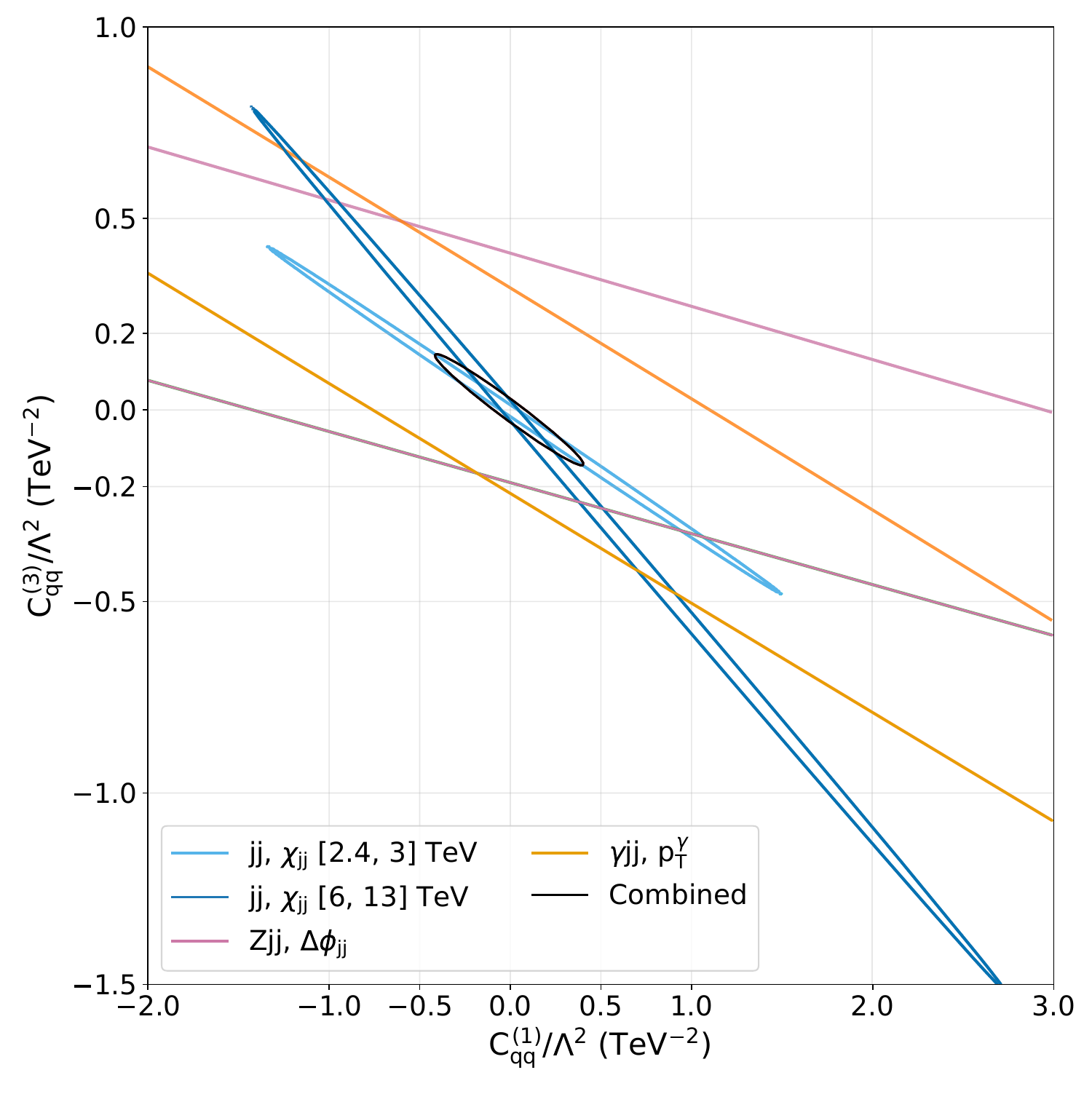}
   \includegraphics[width=.49\textwidth]{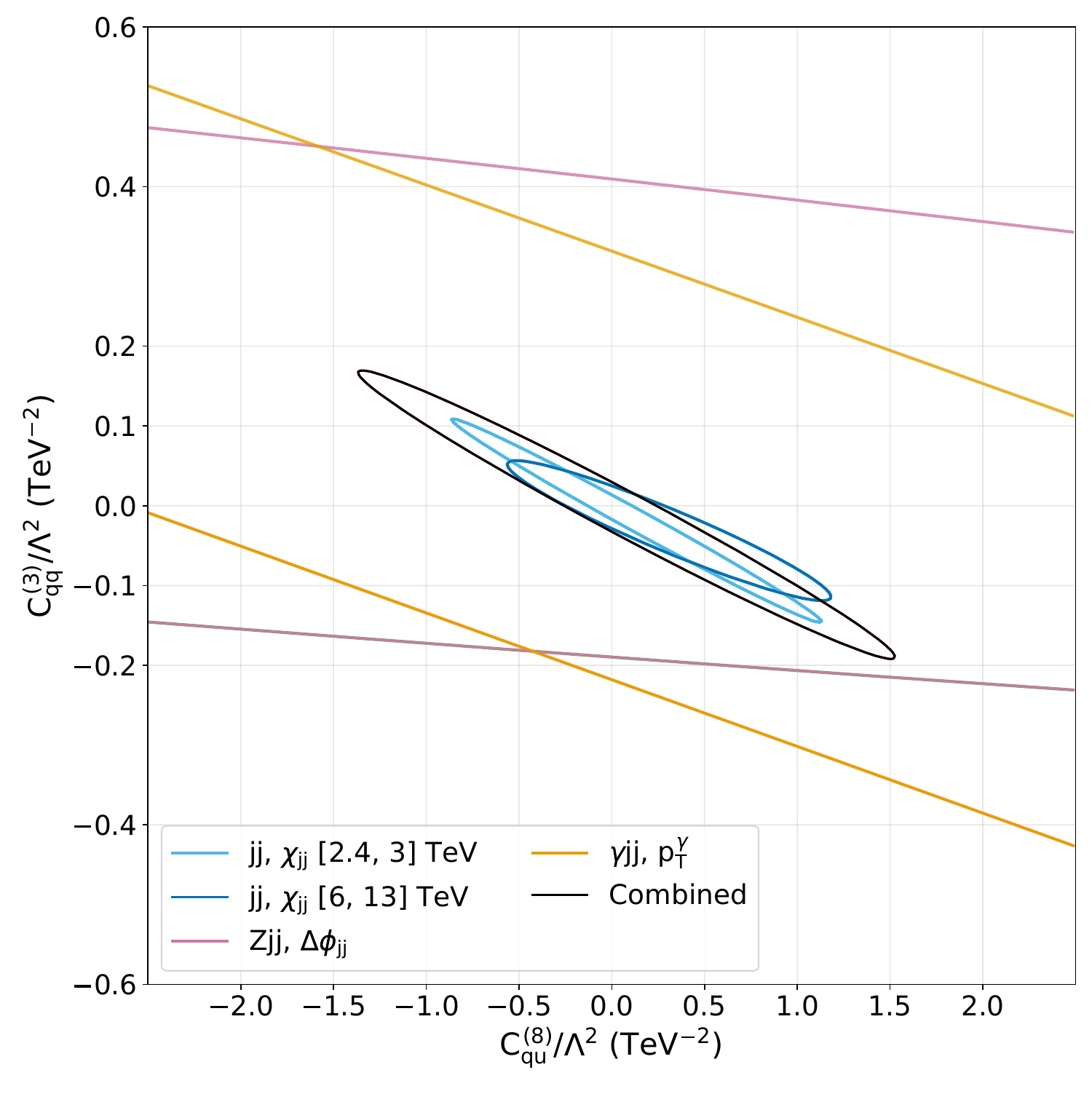}
\end{figure*}

\begin{figure*}   
   \caption{\small{Individual ({\it solid}) and combined marginalised ({\it dashed}) contours in the $C_{qq}^{(1)}$ {\it vs} $C_{qq}^{(3)}$ ({\it left}) and $C_{qu}^{(8)}$ {\it vs} $C_{qq}^{(3)}$ ({\it right}) planes, when only $O_{qq}^{(1)}$, $O_{qq}^{(3)}$, $O_{uu}$ and $O_{dd}$ are included in the fit and only the predictions from $Z+$jets, $\gamma +$jets and the two $M_{jj}$ regions in multijet are used. Note that the ranges are different on the two axes. Individual contours are obtained by setting all the coefficients that are not plotted to 0, while for the marginalised ones they are set to the values that minimise the $\chi^2$ along their directions}} \label{fig:red_contours}
   \includegraphics[width=.49\textwidth]{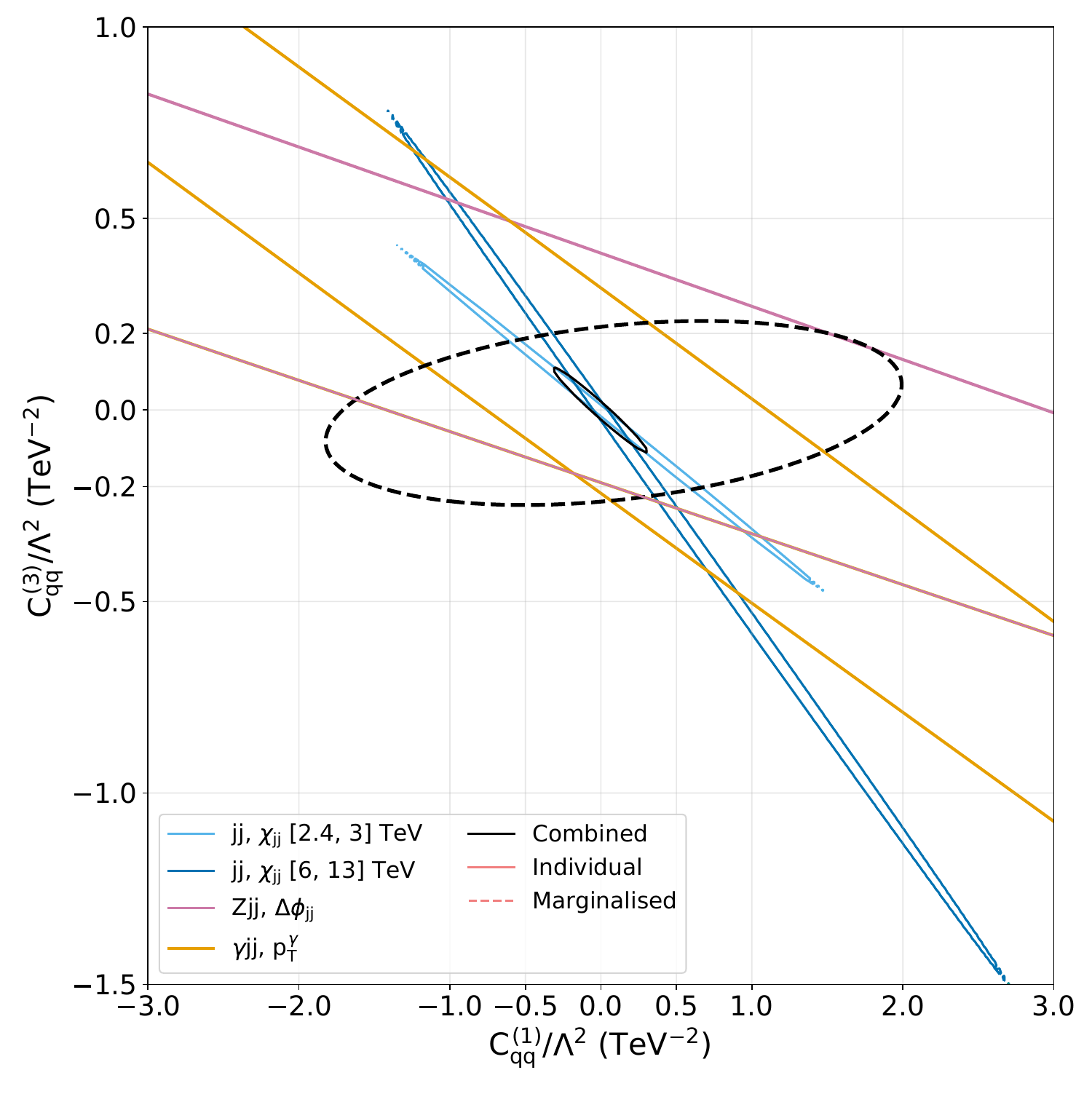}
   \includegraphics[width=.49\textwidth]{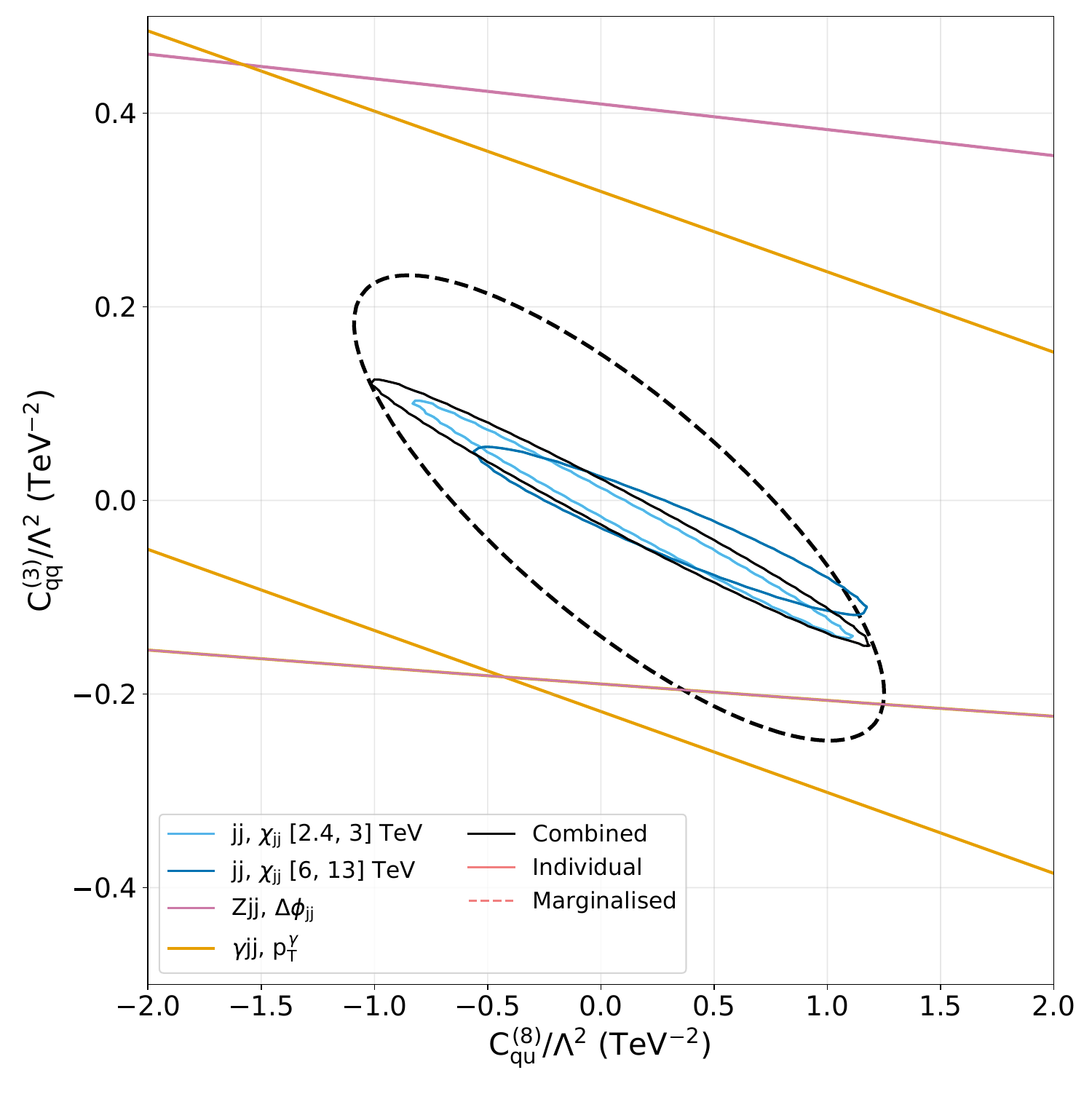}
\end{figure*}

\section{\label{sec:bounds}Individual and marginalised bounds, and squares}
Individual limits on all the 4LQ operators at 95\% Confidence Level (CL) are shown in Fig. \ref{fig:ind_limits}, for the processes and observables described before; the numerical values are listed in Appendix \ref{app:lim}. The combined constraints from all the processes are also reported: they are similar to the ones from multijet. The marginalised bounds are not shown for the single operators, as ten different distribution shapes would be needed for each process, with respect to SM and experimental uncertainties, to obtain meaningful results. This issue could be bypassed by including more variables, even though the bin-per-bin correlations between them are not always available, or by finding differential observables that display larger shape variations amongst the operators. A summary of the experimental datasets used to obtain these limits can be found in Table \ref{tab:4lq_dataset}.

By comparing the interference and partial $\mathcal{O}(1/\Lambda^4)$ cross sections for $O_{qq}^{(3)}$ in Table \ref{tab:all_xsects}, it is possible to observe that the dimension-6 squared contribution is at least as large as the interference one, at the coefficient values obtained in the limits, for all the processes; the only exception is multijet production in the [2.4, 3] TeV $M_{jj}$ region, that provides the most accurate measurements with uncertainties of the order of 10\%. 
As a matter of fact, the individual limits on $C_{qq}^{(3)}$ from the low-$M_{jj}$ dijet measurements are of the order of $2 \cdot 10^{-2}$, and give an interference cross section of the order of 0.36 pb and a squared contribution of 0.04 pb. For coefficient values close to the limits, the dimension-6 squared contribution to the $O_{qq}^{(3)}$ cross section is then only of the order of 10\% of the interference one with the SM, and has little impact on the constraints. On the contrary the high-$M_{jj}$ measurement, for example, yields constraints around $3 \cdot 10^{-2}$, for which the interference cross section would be 2.1 fb and the squared one 1.7 fb: even if this region is probing higher energies, the relative precision is weaker.
Our results confirm that high precision is required to get SMEFT contraints without significant doubts about the validity of the expansion.
For other operators like $O_{ud}^{(1)}$, $O_{qu}^{(1)}$ and $O_{qd}^{(1)}$, that in some cases do not interfere with the SM at all or do it only in some subleading subprocesses, the quadratic contributions will always dominate. It may be worth reminding that the total $\mathcal{O}(1/\Lambda^4)$ term would also include the interference among dimension-8 operators and the SM, which could either increase or decrease significantly the squared contribution.
The relative size of the interference and of the dimension-6 squared contributions hints that the validity of the SMEFT is not guaranteed. A similar conclusion can be reached by comparing the bounds on $\Lambda$ and the energies investigated in the processes, with the hypothesis $C_i\approx 1$: both the scale probed by most of the considered processes and the cutoff scale are around or below 1 TeV. The only exception is multijet, with the best constraints around 5 TeV: this is reasonable for the lowest invariant-mass bin, but again borderline for the highest one.
Fig. \ref{fig:ind_limits} also shows the individual bounds on $C_{qq}^{(3)}/\Lambda^2$ obtained with the inclusion of the squared contributions, from most of the processes mentioned above. The largest improvement in the limits, compared to the interference level, can be observed for $\gamma+$jets, which therefore drives the correspondent combined constraints.

For multijet production, the limits from both the considered $M_{jj}$ intervals are shown. In the [2.4, 3] TeV one, the $\chi^2$ denominator in Eq. \eqref{chisq} is dominated by the SM uncertainties, while in the [6, 13] TeV region the experimental errors are larger. When the interference uncertainty is included as in Eq. \eqref{chisq_2}, the individual limits worsen by factors 1.3 to 1.7. As explained in Sec. \ref{sec:framework}, though, this approach does not incorporate the full correlations among the theoretical uncertainties.

In the $Z+$jets production case, the $\chi^2$ denominator is driven by the experimental uncertainties, that also include the MC-generator choice ones \cite{Atlas:2020zjj}. If the error sources on the interference are also partially considered, the limits become 1.2 to 1.5 times larger.

For $W+$jets, the bounds are obtained by comparing the 4LQ $\mathcal{O}(1/\Lambda^2)$ terms against the LO SM that we generated, and the experimental data is assumed to follow the same distribution, but with different errors. The inclusion of theoretical uncertainties on the interference worsens the bounds by factors between 1.4 and 2.

In $\gamma+$jets production, the absolute uncertainties in the $\chi^2$ denominator are dominated by the SM. With the addition of the interference uncertainties, the limits worsen by factors between 1.2 and 2.3 for all the operators.

For each operator, the most competitive limits come from multijet production, and particularly from the $2.4<M_{jj}<3$ TeV region, where the experimental uncertainties are lower. It is worth reminding the reader, though, that the total experimental cross section is not available for this process, and that the total LO SM one was used instead. $O_{ud}^{(1)}$, $O_{qu}^{(1)}$ and $O_{qd}^{(1)}$, that do not interference with the SM QCD in dijet production and with the main SM $\gamma jj$ channel, are basically unconstrained already at individual level. The most constrained operator is $O_{qq}^{(3)}$, as its contributions are the largest for every process if all coefficients are the same.

The directions in the coefficient space that are best constrained by each process described above are listed in Tab. \ref{tab:constr_dirs}. As discussed in Sec. \ref{sec:framework}, these are not the only vectors that the data could probe, but all the others show eigenvalues that are at least one order of magnitude smaller than the best one, so they would require much larger experimental and theoretical precisions to be investigated. Also, when uncertainties are computed for these other eigenvalues, they all seem to be compatible with zero and thus associated to flat directions. The constraints along them would also be much weaker than those of the best directions, and would therefore raise many questions about the SMEFT validity.

All these most-constrained directions are closer to $C_{qq}^{(3)}/\Lambda^2$ than to the other coefficients, meaning that the respective operator will be the most constrained even in the combined marginalised scenario. Also the limits from $b+$jets production are shown in Fig. \ref{fig:ind_limits} and they are much weaker than for the other processes: this is because of the large scale uncertainties and the low-$p_T$ ranges imposed by the tagging algorithms, that do not help reduce the SM effect. It can be seen from Table \ref{tab:constr_dirs}, though, that the application of the tagging strategies increases the sensitivity to some operators, like $C_{dd}$, compared to the other multijet scenarios. The most constrained line from the sum of all the $\chi^2$ polynomials in the table is very close to the one in the $2.4<M_{jj}<3$ TeV multijet region, as expected from the large eigenvalue of the latter.

We check that different observables for the same processes are probing the same directions in coefficient space, with variations of at most a few percent in the largest components. The same happens even if the variables are plotted over the direct- and fragmentation-enriched regions for $\gamma +$jets. For multijet production, moving to higher energies slightly increases the sensitivity to $O_{qq}^{(1)}$ and $O_{uu}$, compared to $O_{qq}^{(3)}$. Moreover, multijets are able to constrain $O_{qq}^{(1)}$ twice more strongly than $Z+$jets.

We also check that switching between the $\chi^2$ Eqs. \eqref{chisq} and \eqref{chisq_2} to include the linear-term uncertainties does not significantly alter the constrained directions.

The individual contours in the $C_{qq}^{(1)}$ {\it vs} $C_{qq}^{(3)}$ and $C_{qu}^{(8)}$ {\it vs} $C_{qq}^{(3)}$ planes are shown in Fig. \ref{fig:ind_contours}. It can be seen that the most stringent bounds come from the combination of the two multijet $M_{jj}$ regions, that are both included at the same time as the correlations among their bins, reported in \cite{2j_2018}, are much smaller than the ones inside the same ranges. The $Z+$jets and $W+$jets ellipses point along similar directions, as seen in Table \ref{tab:constr_dirs}, and the same happens for $\gamma+$jets and multijets in the [2.4, 3] TeV region. Despite the larger cross section, $W+$jets has a lower constraining power than $Z+$jets because of its looser cuts in energy and the larger uncertainties in the LO SM sample that we generated. The latter could be reduced by going to NLO, but similar pole-cancellation issues as the ones discussed in \cite{cw_paper} for $Zjj$ would have to be tackled. Moreover, more stringent cuts on $M_{jj}$ and the $p_T$ of the jets would increase both the linear and quadratic effects over the SM. 

We then repeat the fit, this time including only the four most constrained operators $O_{qq}^{(1)}$, $O_{qq}^{(3)}$, $O_{uu}$ and $O_{qu}^{(8)}$ and using exclusively the $\chi_{jj}$, $\Delta\phi_{jj}$ and $p_T^\gamma$ differential distributions from the two $M_{jj}$ regions in multijets, $Z+$jets and $\gamma +$jets respectively, since they are the most constraining processes. The best-constrained direction in the coefficient space in this case is the same as in the last column of Table \ref{tab:constr_dirs}, and is associated to the eigenvalue $\lambda = 1.3\cdot 10^5 \pm 50\%$; all the other three eigenvectors seem again to be related to flat directions.
Both the individual and combined marginalised contours for this scenario are shown in Fig. \ref{fig:red_contours} for $C_{qq}^{(1)}$ and $C_{qu}^{(8)}$ {\it vs} $C_{qq}^{(3)}$. The marginalised ellipses are tilted towards the $C_{qq}^{(3)}/\Lambda^2$ axis, as in the individual case, and the associated limits on this coefficient are better by at least a factor $\sim 5$ compared to the ones on the other three included operators. The weaker constraints on the latter make them once again more vulnerable to issues on the validity of the EFT expansion. The marginalised limits in the $C_{qu}^{(8)}$ {\it vs} $C_{qq}^{(3)}$ plane are better than the ones in $C_{qq}^{(1)}$ {\it vs} $C_{qq}^{(3)}$, even though the combined individual bounds are worse; the marginalised contour in the second plane becomes more stringent than in the first one if $C_{uu}$ is set to zero. The same happens if $C_{uu}$ and $C_{qq}^{(1)}$ are swapped in the previous sentence. This suggests that $O_{qq}^{(1)}$ and $O_{uu}$ might partially cancel each other along the main direction probed by the combination of the datasets included in this fit.

\section{Conclusions}
In this paper we showed predictions for the interference among the SM and the 4LQ operators to different processes. We considered multijet as well as jet production in association with EW bosons, like $Z+$jets, $W+$jets and $\gamma +$jets, and different phase-space regions were investigated for some of them. We also simulated a $b$- and $c$-tagging algorithm and applied it to multijets. In each case, we asked for multiple jet multiplicities at parton level, that were then matched, merged and showered. These were expected to have the largest effect on both cross section and distribution shapes for the considered processes, as they involve jets, and we have indeed observed significant differences compared to the partonic level.

For each process and phase-space region, we checked which observables provide the best individual bounds on the 4LQ operators at linear order, and saw that $O_{qq}^{(3)}$ is always the most constrained as its contribution to each process is the largest, if all coefficients are equal. This is partially an effect of the normalisation chosen for this operator: if the $\tau^I = \sigma^I/2$ matrices were used instead of the Pauli ones, as in \cite{2j_SMEFT}, it would induce similar or smaller effects than $O_{qq}^{(1)}$ and $O_{uu}$ in the studied processes. It goes without saying that physics should not depend on the nomalisation: the assumption of some UV models that can run into the 4LQ operators would remove this issue through the matching procedure. Furthermore, depending on the model, this might allow to neglect some of the operators, as the renormalisation group would not mix all of them. Even when keeping model independence, though, using the $\tau^I$ matrices would introduce a factor $1/4$ in front of $O_{qq}^{(3)}$ and simplify the comparison with the other operators, especially when computing the constrained directions and the $\mathcal{O}(1/\Lambda^4)$ estimate.

We also proposed a way to include theoretical uncertainties on the interference distributions in the limit computation, at least partially as their correlations are not always available. 

We computed the directions in the coefficient space that each process and observable can probe and saw that the most constrained ones are close to each other, in particular for the low-$M_{jj}$ multijet region and $\gamma +$jets on one side and $W,Z+$jets on the other. The direction probed by $b+$jets is quite singular, but its constraining power is not competitive with the others. Even the $\gamma, Z, W+$jets processes have low eigenvalues compared to multijets. These directions mainly use the information from the total cross sections; all the other ones, that would include more details about the distribution shapes, seem to be out of the reach of current measurements, as they would need much more precision in the predictions and in particular in the SM ones, where the uncertainties are larger than the experimental ones and MC generators do not agree on the results. As a consequence, only one direction is truly constrained.

For this reason, all the variables that we checked from the same process can only probe very close directions in the coefficient space with different strengths, so we picked the most constraining observable for each of them to obtain marginalised limits on the operators. The main axes of the contours are close to the $( C_{qq}^{(3)} + C_{qq}^{(1)}/2 + C_{uu}/2 )/\Lambda^2$ one every time these operators are contributing, with the strongest bounds coming from multijet production. For all the processes, we tested many of the common variables and found little shape differences between the different operators, compared to the uncertainties. Therefore, new out-of-the-box observables should be tried for separating them.

Due to the fact that the constraints are relatively loose, we expect the next largest correction to be due to the next-order term in the EFT expansion, unless both the SM predictions and the measurements will become significantly more accurate. While RGE effects have not been considered, we do not expect those to be large, as the ranges of scales probed in the various processes are quite close. In addition, the mixing between operators will not induce new shape effects, since all the operators have similar distribution shapes. However, for other processes than multijet, additional operators to the 4LQ need to be taken into account. Eventually, NLO corrections are expected to yield some shape variation if the $K$-factors are not flat: they could help constrain a second direction, but with a still weaker power that the leading one, and therefore rising more concerns about EFT validity issues. As a result, we would strongly recommend to improve first the precision on the SM and the measurements.

We computed the dimension-6 squared corrections for $O_{qq}^{(3)}$ in each process. Because of the high energies requested in the experimental analyses, these contributions are at least as large as the linear ones for the coefficient values that we obtained in our limits. This holds for all processes, with the only exception of the low-$M_{jj}$ region in multijet production. As expected, there is a sweet spot in the energy range where the NP effects are significant compared to the SM, while the experimental precision is still high enough and EFT uncertainties are under control. The interference among the SM and dimension-8 operators needs to be computed, though, to get a complete estimate of the $\mathcal{O}(1/\Lambda^4)$ correction.

To conclude, further improvements of the constraints on the 4LQ operators request both to obtain more precise SM predictions and to find new observables that are more sensitive to their contributions. We also remind the reader that all the distributions that we employed to obtain the limits and the $\chi^2$ functions can be found in a notebook linked to this paper.

\subparagraph{Acknowledgments}
Computational resources have been provided by the Consortium des \'Equipements de Calcul Intensif (C\'ECI), funded by the Fonds de la Recherche Scientifique de Belgique (F.R.S.-FNRS) under Grant No. 2.5020.11 and by the Walloon Region. MM is a Research Fellow of the F.R.S.-FNRS, through the grant \textit{Aspirant}. We are thankful to G. Durieux, K. Mimasu, H. El Faham, I. Brivio, F. Maltoni and V. Lemaitre for the insightful discussion and suggestions they helped us with during this work, and to W. Shepherd and A. Biek\"otter for the support while reproducing their dijet and $\gamma +$jets results, respectively. Our acknowledgments extend as well to all the other participants of the 2025 HEFT conference, hosted at CERN, who shared with us their feedback about this project and its prospects.

\appendix

\begin{table} 
\caption{\small{Analytical expressions of the differential cross section for some dijet subprocesses, in the SM QCD and its interferences with the 4LQ operators, as functions of $\chi_{jj}$. PDF and PS effects are not included. Each interference term has a prefactor $4 C_i \alpha_S^{}/(9\Lambda^2)$, while for the SM ones it is $2\alpha_S^2 \pi /(9s)$, with $s$ the total energy squared of the process}} \label{tab:jj_feynarts}
\resizebox{.45\textwidth}{!}{
\begin{tabular}{c|c}
   \hline
   \multicolumn{2}{c}{$pp\rightarrow jj$, LO} \\
   Operator & $ d\sigma^{1/\Lambda^2} / d\chi_{jj}$ \\
   \hline
   \multicolumn{2}{c}{$u u \rightarrow u u$, $d d \rightarrow d d$, $\bar{u} \bar{u} \rightarrow \bar{u} \bar{u}$, $\bar{d} \bar{d} \rightarrow \bar{d} \bar{d}$} \\
   \hline
   SM QCD  & $\frac{2(3+2\chi+\chi^2+2\chi^3+3\chi^4)}{3\chi^2 (1+\chi)^2}$ \\
   $O_{qq}^{(1)}$, $O_{qq}^{(3)}$, $O_{uu}$, $O_{dd}$ & -$\frac{1}{\chi}$ \\
   $O_{ud}^{(8)}$ & 0 \\
   $O_{qu}^{(8)}$, $O_{qd}^{(8)}$ & -$\frac{1-\chi+\chi^2}{2 \chi (1+\chi)^2}$ \\
   \hline
   \multicolumn{2}{c}{$u d \rightarrow u d$, $\bar{u} \bar{d} \rightarrow \bar{u} \bar{d}$} \\
   \hline
   SM QCD & $\frac{1+2\chi+2\chi^2}{(1+\chi)^2}$ \\
   $O_{qq}^{(1)}$, $O_{uu}$, $O_{dd}$ & 0 \\
   $O_{qq}^{(3)}$ & -$2\frac{1}{1+\chi}$ \\
   $O_{ud}^{(8)}$ & -$\frac{1}{4}\frac{1}{1+\chi}$ \\
   $O_{qu}^{(8)}$, $O_{qd}^{(8)}$ & -$\frac{\chi^2}{4(1+\chi)^3}$ \\
   \hline
   \multicolumn{2}{c}{$u \bar{d} \rightarrow u \bar{d}$, $d \bar{u} \rightarrow d \bar{u}$} \\
   \hline
   SM QCD & $\frac{1+2\chi+2\chi^2}{(1+\chi)^2}$ \\
   $O_{qq}^{(1)}$, $O_{uu}$, $O_{dd}$ & 0 \\
   $O_{qq}^{(3)}$ & -$2\frac{\chi^2}{(1+\chi)^3}$ \\
   $O_{ud}^{(8)}$ & -$\frac{1}{4}\frac{\chi^2}{(1+\chi)^3}$ \\
   $O_{qu}^{(8)}$, $O_{qd}^{(8)}$ & -$\frac{1}{4(1+\chi)}$ \\
   \hline
   \multicolumn{2}{c}{$u \bar{u} \rightarrow u \bar{u}$, $d \bar{d} \rightarrow d \bar{d}$} \\
   \hline
   SM QCD & $\frac{27+12\chi+29\chi^2+2\chi^3+6\chi^4}{3(1+\chi)^4}$\\
   $O_{qq}^{(1)}$, $O_{qq}^{(3)}$, $O_{uu}$, $O_{dd}$ & -$\frac{\chi^3}{(1+\chi)^4}$ \\
   $O_{ud}^{(8)}$ & 0 \\
   $O_{qu}^{(8)}$, $O_{qd}^{(8)}$ & -$ \frac{\chi (3+3\chi+\chi^2)}{2(1+\chi)^4}$ \\
   \hline
\end{tabular}
}
\end{table}

\begin{figure}
   \caption{\small{Shapes of the differential cross section with respect to $\chi_{jj}$ for the SM and its interference with the 4LQ operators, for some subprocesses to dijet production. PDF and PS effects are not included}} \label{fig:dxsdChi_part2}
   \includegraphics[width=0.49\textwidth]{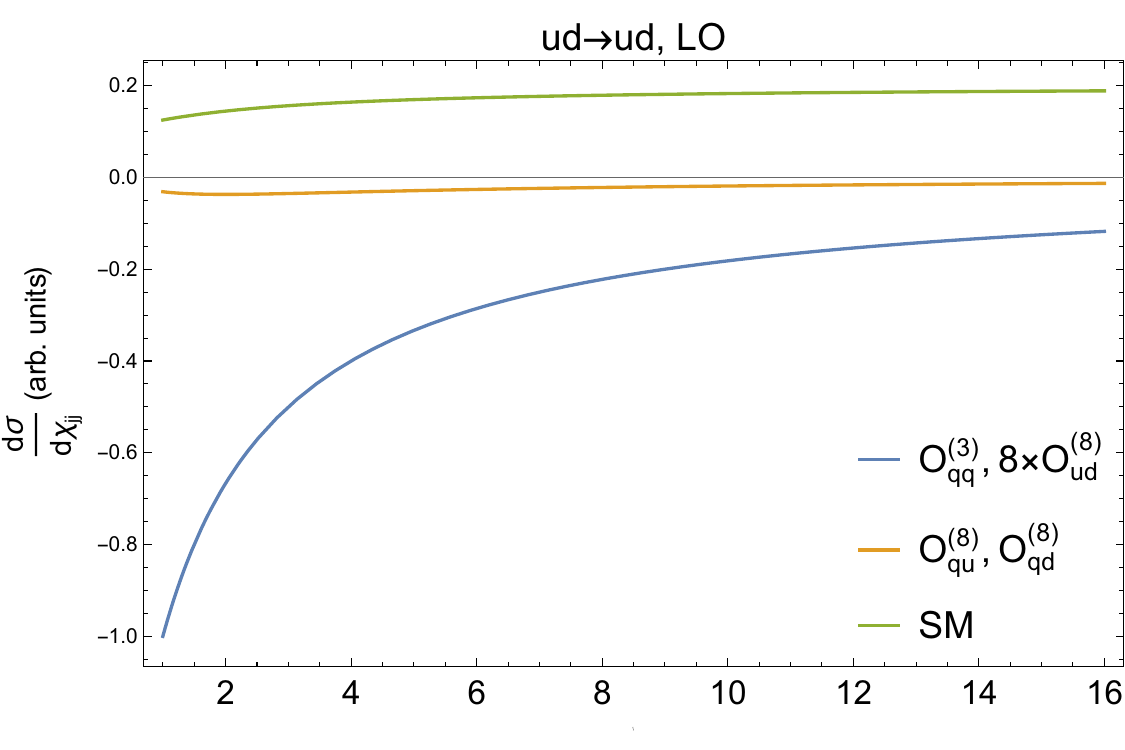}
   \includegraphics[width=0.49\textwidth]{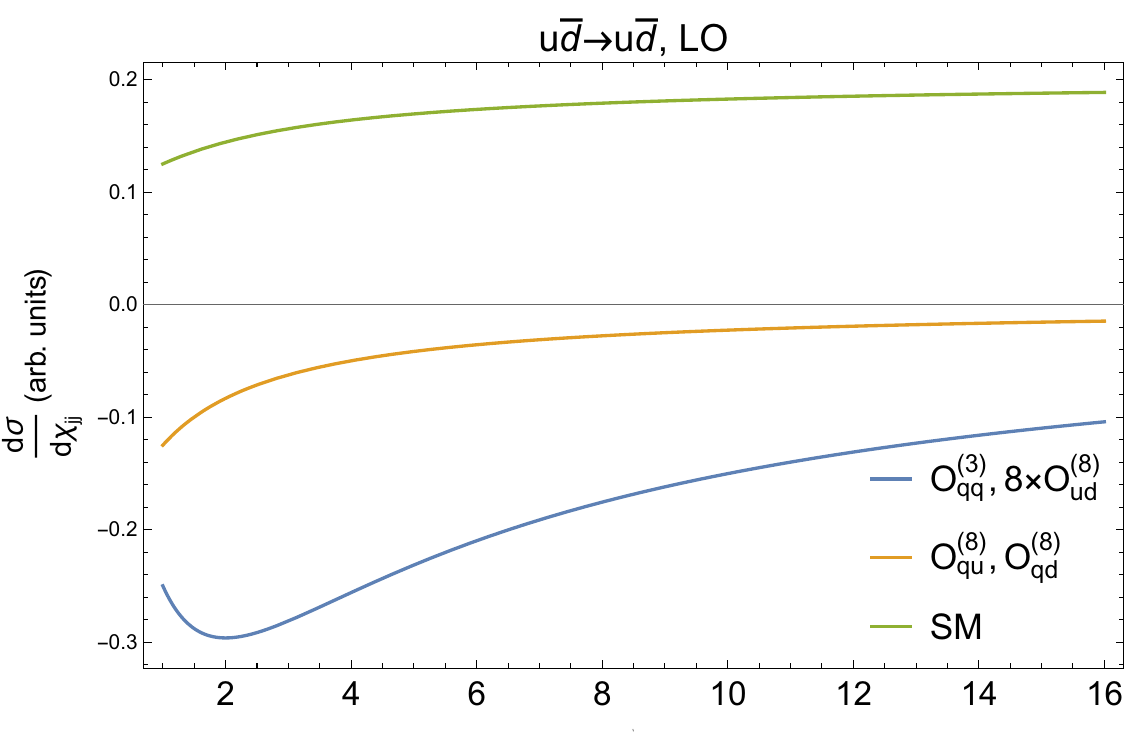}
   \includegraphics[width=0.49\textwidth]{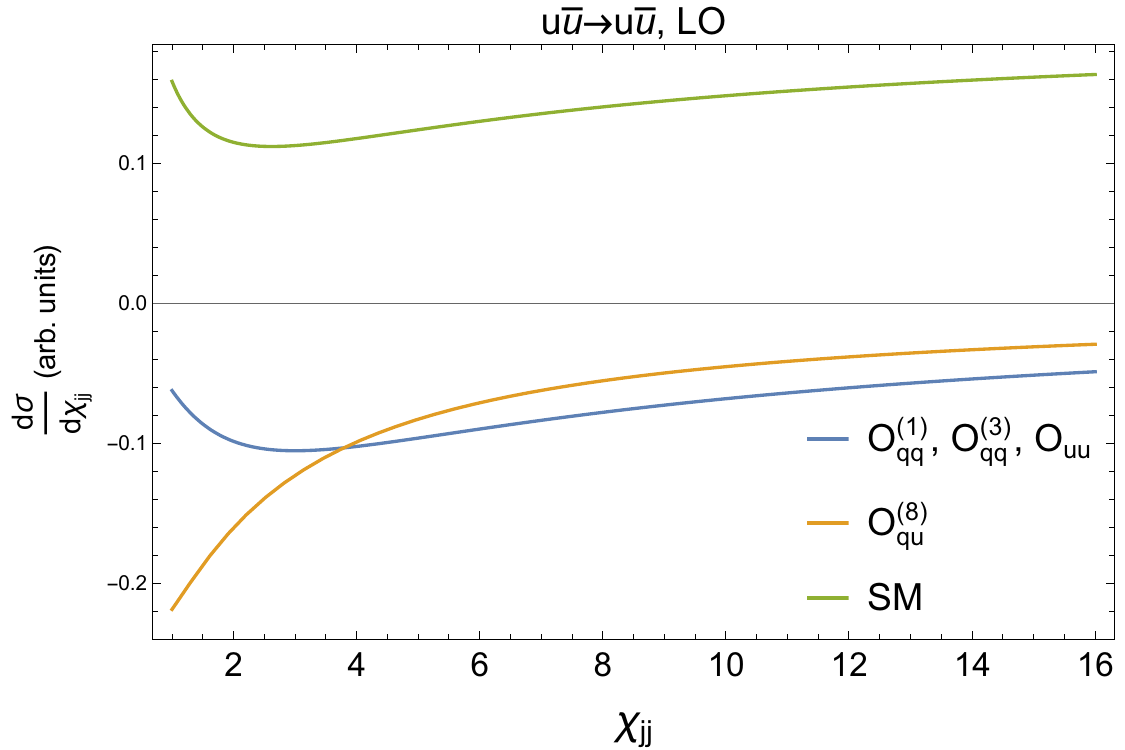}
\end{figure}

\begin{table}
\caption{\small{Cross sections, in pb, for multijet production when at least one $b$-jet, at least one $c$-jet and at least two $b$-jets are tagged, at LO matched to PS. The SM and 4LQ interference values are reported. The scale-variation uncertainties are shown}} \label{tab:tag_xsects}
\begin{tabular}{c|ccc}
\hline
\multicolumn{4}{c}{$pp \rightarrow$ jets} \\ 
 & $b+$jets & $c+$jets & $bb+$jets \\ \hline
SM & 2.9$\cdot 10^6 {}^{+69\%}_{-44\%}$ & 1.9$\cdot 10^6 {}^{+57\%}_{-42\%}$ & 2.5$\cdot 10^6 {}^{+54\%}_{-35\%}$ \\
$O_{qq}^{(1)}$ & -8$\cdot 10^2 {}^{+38\%}_{-62\%}$ & -4$\cdot 10^2 {}^{+40\%}_{-50\%}$ & 1$\cdot 10^1 {}^{+400\%}_{-400\%}$ \\
$O_{qq}^{(3)}$ & -2.3$\cdot 10^3 {}^{+43\%}_{-65\%}$ & -1.4$\cdot 10^3 {}^{+43\%}_{-57\%}$ & 7$\cdot 10^1 {}^{+142\%}_{-142\%}$ \\
$O_{uu}$ & -1.6$\cdot 10^2 {}^{+44\%}_{-63\%}$ & -3.7$\cdot 10^2 {}^{+41\%}_{-54\%}$ & -2${}^{+100\%}_{-100\%}$ \\
$O_{dd}$ & -6$\cdot 10^2 {}^{+50\%}_{-67\%}$ & -3$\cdot 10^1 {}^{+67\%}_{-67\%}$ & 1$\cdot 10^1 {}^{+400\%}_{-400\%}$ \\
$O_{ud}^{(8)}$ & -1.9$\cdot 10^2 {}^{+42\%}_{-63\%}$ & -1.3$\cdot 10^2 {}^{+38\%}_{-62\%}$ & 7${}^{+114\%}_{-114\%}$ \\
$O_{qu}^{(8)}$ & -2.6$\cdot 10^2 {}^{+42\%}_{-62\%}$ & -2.9$\cdot 10^2 {}^{+41\%}_{-59\%}$ & 6 ${}^{+150\%}_{-150\%}$ \\
$O_{qd}^{(8)}$ & -5$\cdot 10^2 {}^{+20\%}_{-60\%}$ & -1.3$\cdot 10^2 {}^{+46\%}_{-62\%}$ & 1$\cdot 10^1 {}^{+300\%}_{-300\%}$ \\
\hline
\end{tabular}
\end{table}

\section{\label{app:dijet}Dijet differential cross sections}
The analytical expressions for the dijet differential cross sections, with respect to the cosine of the scattering angle $\theta^*$ in the CoM frame and to $\chi_{jj}$, are listed in Table \ref{tab:jj_feynarts}, split for different subprocesses. The shapes for some of them are shown in Figures \ref{fig:dxsdChi} and \ref{fig:dxsdChi_part2}. Even if the subprocesses present different trends, the one from $uu\rightarrow uu$ still dominates when they are summed together, as it can be seen in Fig. \ref{fig:jj_chijj}.

\section{\label{app:jj_tag}Multijet with flavour-tagging}
Table \ref{tab:tag_xsects} shows the cross sections for multijet production with the employment of simulated flavour-tagging algorithms to identify at least one $b$-jet, one $c$-jet or two $b$-jets. More details about the implementation can be found in Sec. \ref{sec:framework}. The values for the SM and its interference with the 4LQ operators are shown, at LO matched with MLM and PS.

\section{\label{app:ajj}$\gamma+$jets cross sections in the direct- and fragmentation-enriched regions}
As stated in Sec. \ref{sec:ajj}, the experimental analysis \cite{Atlas:2020ajj} for $\gamma+$jets productions defines, together with an inclusive region, a direct-enriched ($p_T^\gamma > p_T^{j1}$) and a fragmentation-enriched ($p_T^\gamma < p_T^{j2}$) ones. The total cross sections for these last two are listed in Table \ref{tab:ajj_frag_dir}. It can be observed that the direct-region cross section is larger than the fragment-region one in the SM and experimental measurments, but the opposite holds for almost all the interference terms (in absolute value).

\section{\label{app:lim}Numerical values of the individual limits}
The numerical values for the individual limits shown in Fig. \ref{fig:ind_limits} on the 4LQ coefficients are reported in Table \ref{tab:ind_limits}, from each process included in this study.

\begin{table}[ht]
\caption{\small{LO cross sections, in fb, for the SM and its interference with the 4LQ operators in $\gamma$+jets production, in the fragmented- ({\it left}) and direct-enriched ({\it right}) phase-space regions described in the text. The experimental measurements are also reported. For these and for the SM, the cumulative uncertainty is shown, while for the interference terms the first source is numerical and the following ones come from scale variations}} \label{tab:ajj_frag_dir}
\begin{tabular}{c|cc}
\hline
\multicolumn{3}{c}{$pp \rightarrow \gamma +$jets} \\ 
 & $p_T^\gamma < p_T^{j2}$ & $p_T^\gamma > p_T^{j1}$ \\ \hline 
Exp & 3.5$\cdot 10^3 \pm 4\%$ & 1.17$\cdot 10^4 \pm 6\%$ \\ 
SM & 2.6$\cdot 10^3 \pm 30\%$ & 1.3$\cdot 10^4 \pm 31\%$ \\
$O_{qq}^{(1)}$ & -2$\cdot 10^2 \pm 0.3\%^{+15\%}_{-20\%}$ & -4.7$\cdot 10^1 \pm 0.9\%^{+17\%}_{-21\%}$ \\
$O_{qq}^{(3)}$ & -8.9$\cdot 10^2 \pm 0.2\%^{+10\%}_{-12\%}$ & -3.0$\cdot 10^2 \pm 0.5\%^{+10\%}_{-10\%}$ \\
$O_{uu}$ & -1.7$\cdot 10^2 \pm 0.2\%^{+12\%}_{-17\%}$ & -5.5$\cdot 10^1 \pm 0.4\%^{+9\%}_{-11\%}$ \\
$O_{dd}$ & -1.6$\cdot 10^1 \pm 0.3\%^{+13\%}_{-19\%}$ & -6.6$\pm 0.5\%^{+11\%}_{-12\%}$ \\
$O_{ud}^{(1)}$ & 7.1$\pm0.3\%^{+28\%}_{-18\%}$ & 3.1$\pm 0.5\%^{+23\%}_{-16\%}$ \\
$O_{ud}^{(8)}$ & -4.7$\cdot 10^1 \pm0.3\%^{+9\%}_{-15\%}$ & -2.6$\cdot 10^1 \pm 0.4\%^{+7\%}_{-12\%}$ \\
$O_{qu}^{(1)}$ & -5.4$\pm 0.6\%^{+30\%}_{-35\%}$ & -2.5$\pm 0.8\%^{+24\%}_{-32\%}$ \\
$O_{qu}^{(8)}$ & -9$\cdot 10^1 \pm 0.3\%^{+19\%}_{-26\%}$ & -4.1$\cdot 10^1 \pm 0.5\%^{+14\%}_{-20\%}$ \\
$O_{qd}^{(1)}$ & 9$\cdot 10^{-1} \pm 1.0\%^{+78\%}_{-67\%}$ & 3$\cdot 10^{-1} \pm 2\%^{+67\%}_{-67\%}$ \\
$O_{qd}^{(8)}$ & -3.3$\cdot 10^1 \pm 0.3\%^{+18\%}_{-27\%}$ & -1.6$\cdot 10^1 \pm 0.5\%^{+13\%}_{-19\%}$ \\
\hline
\end{tabular}
\end{table}

\begin{table*}
\caption{\small{Numerical values of the 95\% CL individual limits on the 4LQ coefficients $C_i/\Lambda^2$, in TeV${}^{-2}$, from the various processes included in this study. The bounds at square level are also shown for $C_{qq}^{(3)}$ for some processes. All the numbers in the table are graphically represented in Fig. \ref{fig:ind_limits}}} \label{tab:ind_limits}
\begin{tabular}{c|ccccccc}
   \hline
    & \makecell{$jj$, $\chi_{jj}$ \\ $[2.4, 3]$ TeV} & \makecell{$jj$, $\chi_{jj}$ \\ $[6, 13]$ TeV} & $bj$, $p_T^b$ & $Zjj$, $\Delta \phi_{jj}$ & $\gamma jj$, $p_T^\gamma$ & $Wjj$, $\Delta \phi_{jj}$ & Combined \\ \hline
   $C_{qq}^{(1)}$ & [-5.5, 4.5]$\cdot 10^{-2}$ & [-5.7, 5,1]$\cdot 10^{-2}$ & [-5.87, 5.8]$\cdot 10^1$ & [-1.5, 3.1] & [-7.9, 11.3]$\cdot 10^{-1}$ & [-1.3, 1.3]$\cdot 10^1$ & [-9.3, 8.4]$\cdot 10^{-2}$ \\
   $C_{qq}^{(3)}$ & [-1.8, 1.4]$\cdot 10^{-2}$ & [-3.2, 2.8]$\cdot 10^{-2}$ & [-1.8, 1.8]$\cdot 10^1$ & [-2.0, 4.2]$\cdot 10^{-1}$ & [-2.3, 3.3]$\cdot 10^{-1}$ & [-2.8, 2.8] & [-3.3, 3.0]$\cdot 10^{-2}$ \\
   $C_{qq}^{(3)} \hspace{1mm} \mathcal{O}(1/\Lambda^4)$ & [-1.7, 1.5]$\cdot 10^{-2}$ & [-1.5, 2.9]$\cdot 10^{-2}$ & - & [-1.9, 6.3]$\cdot 10^{-1}$ & [-4.2, 4.3]$\cdot 10^{-3}$ & [-1.5, 2.6] & [-5.6, 5.6]$\cdot 10^{-3}$ \\
   $C_{uu}$ & [-7.3, 5.9]$\cdot 10^{-2}$ & [-7.9, 7.2]$\cdot 10^{-2}$ & [-9.3, 9.4]$\cdot 10^2$ & [-2.5, 3.8]$\cdot 10^1$ & [-7.9, 11.4]$\cdot 10^{-1}$ & - & [-1.2, 1.1]$\cdot 10^{-1}$ \\
   $C_{dd}$ & [-3.3, 2.7]$\cdot 10^{-1}$ & [-1.3, 1.2] & [-8.6, 8.6]$\cdot 10^1$ & [-2.9, 4.3]$\cdot 10^2$ & [-2.6, 3.8]$\cdot 10^1$ & - & [-6.5, 5.8]$\cdot 10^{-1}$ \\
   $C_{ud}^{(1)}$ & - & - & - & [-10.0, 7.7]$\cdot 10^2$ & [-3.5, 2.4]$\cdot 10^1$ & - & [-6.2, 5.1]$\cdot 10^1$ \\
   $C_{ud}^{(8)}$ & [-3.1, 2.5]$\cdot 10^{-1}$ & [-6.6, 6.0]$\cdot 10^{-1}$ & [-2.3, 2.3]$\cdot 10^2$ & [-1.0, 1.2]$\cdot 10^2$ & [-4.2, 6.1] & - & [-5.9, 5.3]$\cdot 10^{-1}$\\
   $C_{qu}^{(1)}$ & - & - & - & [-1.8, 2.4]$\cdot 10^2$ & [-6.6, 9.8]$\cdot 10^1$ & [-1.5, 1.5]$\cdot 10^3$ & [-1.3, 1.7]$\cdot 10^2$\\
   $C_{qu}^{(8)}$ & [-1.4, 1.1]$\cdot 10^{-1}$ & [-2.8, 3.1]$\cdot 10^{-1}$ & [-2.1, 2.1]$\cdot 10^2$ & [-1.1, 1.4]$\cdot 10^1$ & [-2.7, 3.9] & [-8.9, 8.9]$\cdot 10^1$ & [-2.6, 2.4]$\cdot 10^{-1}$ \\
   $C_{qd}^{(1)}$ & - & - & - & [-8.8, 6.4]$\cdot 10^2$ & [-2.9, 2.0]$\cdot 10^2$ & [-8.7, 8.7]$\cdot 10^3$ & [-5.0, 4.1]$\cdot 10^2$ \\
   $C_{qd}^{(8)}$ & -[2.4, 2.0]$\cdot 10^{-1}$ & [-8.2, 9.2]$\cdot 10^{-1}$ & [-1.1, 1.1]$\cdot 10^2$ & [-2.0, 2.8]$\cdot 10^1$ & [-1.2, 1.8]$\cdot 10^1$ & [-1.1, 1.1]$\cdot 10^2$ & [-4.7, 4.3]$\cdot 10^{-1}$ \\ \hline
\end{tabular}
\end{table*}

\bibliography{refs.bib}
\end{document}